\documentclass[a4paper,11pt,onecolumn,allowfontchangeintitle, accepted=2024-12-10]{quantumarticle} % unpublished,, accepted=2017-05-09
\pdfoutput=1

\usepackage[numbers,sort&compress]{natbib}

\usepackage{hyperref}
\usepackage{xurl} % Allows better URL breaking
\usepackage{url}
\usepackage{doi}
\hypersetup{
    breaklinks=true,
    colorlinks=true,
    linkcolor=blue,
    citecolor=blue,
    urlcolor=blue
}

\usepackage[utf8]{inputenc}
\usepackage{graphicx, float, amsmath, physics}
\usepackage{amssymb,amsfonts,latexsym, amsthm}
\usepackage{subfigure}
\usepackage{enumerate} 
\usepackage{ dsfont } 
\usepackage{tcolorbox}
\usepackage{ragged2e}
\usepackage[a4paper, total={7in, 10in}]{geometry}
\usepackage{tikz}
\usepackage[frozencache,cachedir=minted-cache]{minted}

\usepackage{algorithm}
\usepackage{algpseudocode}

\usepackage{cleveref}
\usepackage{xcolor}

\newcommand{\Dodis}{\mathsf{Dodis \hspace{0.1cm} et \hspace{0.1cm} al.}}

\newcommand{\Toeplitz}{\mathsf{Toeplitz}}
\newcommand{\Trevisan}{\mathsf{Trevisan}}
\newcommand{\Circulant}{\mathsf{Circulant}}
\newcommand{\VonNeumann}{\mathsf{Von \hspace{0.1cm} Neumann}}
\newcommand{\NTT}{\mathsf{NTT}}

\newcommand{\Cryptomite}{\texttt{Cryptomite}}
\def\E{\mathcal{E}}
\def\X{\mathcal{X}}

\def\Z{\mathcal{Z}}

\theoremstyle{definition}
\newtheorem{theorem}{Theorem}

\newtheorem{lemma}{Lemma}
\newtheorem{corollary}{Corollary}

\theoremstyle{definition}
\theoremstyle{definition}
\newtheorem{definition}{Definition}
\theoremstyle{definition}
\theoremstyle{definition}
\theoremstyle{definition}

\usepackage{hyperref}
\usepackage[normalem]{ulem}

\fancyheadoffset{0cm} 

\begin{document}

\title{Cryptomite: \newline A versatile and user-friendly library of randomness extractors} %

\author{Cameron Foreman}
\email{cameron.foreman@quantinuum.com}
\affiliation{Quantinuum, Partnership House, Carlisle Place, London SW1P 1BX, United Kingdom}
\affiliation{Department of Computer Science, University College London, London, United Kingdom}

\author{Richie Yeung}
\email{richie.yeung@quantinuum.com}
\affiliation{Quantinuum, 17 Beaumont Street, Oxford OX1 2NA, United Kingdom}
\affiliation{Department of Computer Science, University of Oxford, Oxford, United Kingdom}

\author{Alec Edgington}
\affiliation{Quantinuum, Terrington House, 13–15 Hills Road, Cambridge CB2 1NL, United Kingdom}

\author{Florian J. Curchod}
\email{florian.curchod@quantinuum.com}
\affiliation{Quantinuum, Terrington House, 13–15 Hills Road, Cambridge CB2 1NL, United Kingdom}

\begin{abstract}
  We present \texttt{Cryptomite}, a Python library of randomness extractor implementations. 
  The library offers a range of two-source, seeded and deterministic randomness extractors, together with parameter calculation modules, making it easy to use and suitable for a variety of applications. 
  We also present theoretical results, including new extractor constructions and improvements to existing extractor parameters.
  The extractor implementations are efficient in practice and tolerate input sizes of up to $2^{40}>10^{12}$ bits. Contrary to alternatives using the fast Fourier transform, we implement convolutions efficiently using the number-theoretic transform to avoid rounding errors, making them well suited to cryptography.
  The algorithms and parameter calculation are described in detail, including illustrative code examples and performance benchmarking.
\end{abstract}

\maketitle

\tableofcontents

\section{Introduction}
\label{sec:Introduction}
Informally, a randomness extractor is a function that outputs \textit{(near-)perfect} randomness (in the sense that it is almost uniformly distributed) when processing an input that is `somewhat' random (in the sense that it may be far from uniformly distributed)\@. 
Randomness extractors play an essential role in many cryptographic applications, for example as exposure-resilient functions, to perform privacy amplification in quantum key distribution or to distil cryptographic randomness from the raw output of an entropy source. 
Beyond cryptography, randomness extractors are a useful tool for many tasks, including the de-randomisation of algorithms or constructing list-decodable error-correcting codes. For an introduction to randomness extraction, see \cite{shaltiel2011introduction}.\\

Despite their usefulness, a major difficulty is to select the appropriate randomness extractor and its parameters for the task at hand -- and more generally, to design, optimise and implement the algorithm without significant expertise and time investment. 
To address this challenge, we have developed \texttt{Cryptomite}, a software library that offers state-of-the-art randomness extractors that are easy to use (see \Cref{sec:showcase}: `Cryptomite: Examples with Code') and efficient (see \Cref{sec:performance}: `Performance'). {Our implementations are also numerically precise, in the sense that the extractor algorithms do not use any floating point arithmetic that can lead to rounding errors. This is accomplished using the number-theoretic transform (NTT) \cite{van2006quantum}, instead of the fast Fourier transform (FFT) \cite{nussbaumer1982fast}, whenever implementing convolutions efficiently.\footnote{{The NTT can be seen as a generalisation of the FFT to finite fields, allowing to perform fast convolutions on integer sequences. The FFT directly performs fast convolutions using floating point numbers and thus precision limited by the amount of memory dedicated to it.}} This is important since uncontrolled errors can lead to implementation vulnerabilities or failures, making our implementations more reliable than alternatives.}\\

The library contains extractors from existing work and our own constructions, along with several improvements to parameters and security proofs.
With the library, we provide explanations, theorems and parameter derivation modules to assist the user and ensure that they make a good extractor choice for their application. 
\texttt{Cryptomite} is made available at \href{https://github.com/CQCL/cryptomite}{https://github.com/CQCL/cryptomite} under a public licence \href{https://github.com/CQCL/cryptomite/blob/main/LICENSE}{https://github.com/CQCL/cryptomite/blob/main/LICENSE} for non-commercial use, and installable through \texttt{pip}, with command $\texttt{pip install cryptomite}$\@. 
In \cite{foreman23stat}, we additionally show how to use the extractors in $\Cryptomite$ to post-process the output of several (quantum) random number generators and study their effect using intense statistical testing.\\

In the following, we give an overview of different types of randomness extractor (\Cref{sec:extractors-at-a-glance}) and define our contributions (\Cref{sec:contributions}). Then, we present the $\Cryptomite$ library (\Cref{sec:randomness-extractor-implementations}) at a high level, describing each extractor and benchmarking the library's performance, as well as providing a simple, visual guide to help a user in selecting the appropriate extractor. After that, we present the library in full detail (\Cref{sec:library-in-detail}), where we give useful theorems and lemmas related to randomness extraction and explain how to calculate all relevant parameters. We finish with code examples (\Cref{sec:showcase}) on how to use $\Cryptomite$ extractors in experimental quantum key distribution and randomness generation, as well as the improvements that are achieved by using extractors of $\Cryptomite$.

\subsection{Randomness extraction at a glance} \label{sec:extractors-at-a-glance}
Randomness extractors transform strings of weakly random numbers (which we call the \textit{source} or \textit{input}), in the sense that their distribution has some min-entropy (the most conservative measure of a random variable's unpredictability, see \Cref{def:min-ent}), into {(near-)perfect} randomness, in the sense that the output's distribution is {(near-)}indistinguishable from uniform (see Definitions \ref{def:classical-secure} and \ref{def:quantum-secure}), as quantified by an extractor error $\epsilon$\@. 
Some extractors require an additional independent string of (near-)perfect random numbers for this transformation, which we call the \textit{seed}\footnote{{Remark that, for useful seeded extractors, the  seed is either short or the extractor is strong (which means that its output is (almost) independent of the seed and therefore the seed is not \textit{consumed} in the process, see \Cref{def:strong-extractor} for the formal statement).}}. {Additionally, we call the seed a \textit{weak seed} if it is not (near-)perfect and has only a min-entropy guarantee.}\\

\begin{figure}[H]
	\centering
	\includegraphics[width=0.6\textwidth]{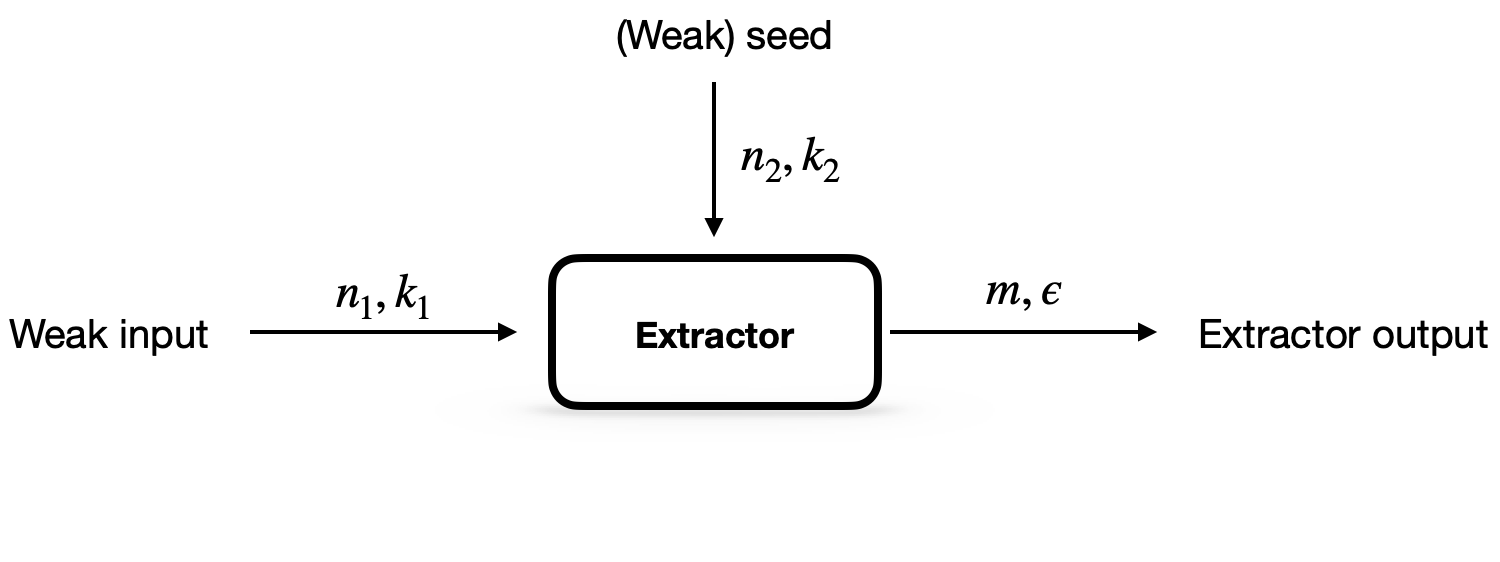}
	\caption{An extractor processes a weak input of $n_1$ bits with min-entropy $k_1$ in order to produce an output of $m$ bits that is $\epsilon$-perfectly random. Seeded and two-source extractors additionally require a {second random bit string called the (weak) seed, of $n_2$ bits and min-entropy $k_2$ (with $k_2=n_2$ for seeded extractors).}\label{fig:Ext_schematics}}
\end{figure}

In the manuscript, we use the notation $n_1, n_2$ to denote the lengths (in bits) of the weakly random extractor input and the (weak) seed respectively. 
We use $k_1, k_2$ (respectively) to denote their min-entropy, $m$ the extractor's output length, $\epsilon$ the extractor error and {$O(\cdot)$} the asymptotic quantities related to the extractor algorithm, for example its computation time. The set-up and variables are illustrated in \Cref{fig:Ext_schematics}\@.

At a high level, randomness extractors can be described by three classes:
\begin{itemize} 
	\item \textbf{Deterministic extractors} are algorithms that process the weak input alone, without additional resources.
    Such extractors are able to extract from a subset of weakly random sources only, but require certain properties of the input distribution beyond just a min-entropy promise -- for example that every bit is generated in an independent and identically distributed (IID) manner. 
    A list of works detailing the different types of input which can be deterministically extracted from can be found in \cite{gabizon2006deterministic} (Subsection `Some related work on randomness extraction')\@. 
	\item \textbf{Seeded extractors} require an additional string of (near-)perfect randomness, called a {seed}, which is generated independently of the extractor weak input. 
    By leveraging this additional independent randomness provided by the seed, they can extract from weak random inputs characterised by min-entropy only. 
	\item \textbf{Multi-source extractors}, sometimes called blenders, are a generalisation of seeded extractors. {They require at least one additional independent weak source of randomness, rather than a single (near-)perfect and independent one.}
    In this work, the only multi-source extractors we consider are \textbf{two-source}, i.e.\ where the user has one addition weak input string only, {which we call the weak seed}. 
\end{itemize}

When using randomness extraction algorithms in real-world protocols, the following features are crucial:
\begin{itemize}
    \item[] \textbf{Implementability and efficiency}: Some randomness extractors only have non-constructive proofs and are therefore theoretical objects that can only be used to derive fundamental results. 
    Moreover, although a randomness extractor may have an explicit construction, this does not always mean that the algorithm can be implemented with a computation time suitable for the application at hand. For example, taking directly the building blocks in \cite{raz2005extractors} would lead to a computation time of $O(n_1^4)$ (for $n_1 \propto n_2$), which significantly limits the maximum input lengths that the implementation can handle. 
    For most applications, a computation time of $O(n_1^2)$ is a requirement, whilst many even require quasi-linear computation time $O(n_1\log(n_1))${ -- such as quantum key distribution. For an excellent discussion and concrete example of the need for quasi-linear computation time, see Appendix E, Section C of \cite{hayashi-tsurumaru} and our examples with code in \Cref{sec:showcase}}. 
    \item[] \textbf{Additional resources}: Seeded and two-source extractors require a seed, which is either near-perfect (seeded extractors) or weak (two-source extractor). In addition to the weakly random input to extract from, this {obliges} the user to have access to, or be able to generate, an independent (weak) seed, which can be difficult to justify (and often not even discussed). {Because of this, it is important to minimise the resources required by the extractor, for example, by extracting with a (near-)perfect seed that is only short \cite{trevisan1999construction, hayashi-tsurumaru} or minimising the entropy requirement on the weak seed \cite{chattopadhyay2022extractors, li2023two}.}
     \item[] \textbf{Security options:} Certain randomness extractor constructions have been shown to fail against a quantum adversary who can store information about {extractor inputs} in quantum states \cite{gavinsky2007exponential}. Because of this, it is important to have the option to choose extractors {that} are \textit{quantum-proof}, for example, in protocols such as quantum key distribution.
     Similarly, in certain protocols the adversary is able to perform actions that may correlate the input and the (weak) seed, therefore breaking the independence condition. This is the case for example in randomness amplification or could happen in quantum key distribution. Consequently, techniques have been developed that allow to weaken the independence condition \cite{ball2022randomness, markov-extractors, dodis2020extracting}.
     \item[] \textbf{Entropy loss}: Entropy loss is the difference between the min-entropy of the extractor input and the length of the extractor output.
     Some randomness extractors are able to extract with entropy loss that is only logarithmic in the inverse extractor error, which is the fundamental minimum \cite{radhakrishnan2000bounds} against both quantum and classical adversaries.
\end{itemize}

In $\Cryptomite$, we implemented several randomness extractors that have a variety of the important features presented above. 
{Our extractors are efficient, with $O(n_1\log n_1)$ or $O(n_1^2\log^2n_1)$ computation time, and have minimal or near-minimal entropy loss. They can extract from various initial resources and come with many options; such as whether to make them quantum-proof or secure when the input and (weak) seed are not fully independent. Additionally, all of our extractors are \textit{strong} (meaning the output is statistically independent of one input), information-theoretically secure (secure against computationally unbounded adversaries) and \textit{universally composable} \cite{canetti2001universally} (enabling secure integration into broader cryptographic protocols).}

\subsection{Our contributions}
\label{sec:contributions}

The contributions of our work are:
\begin{enumerate}
	\item We provide an efficient code implementation of a two-source $\Dodis$ extractor, based on \cite{dodis2004improved} and our previous work \cite{foreman2020practical}\@. It is implemented in quasi-linear computation time {$O(n_1\log n_1)$}\@. {To our knowledge, there exist no alternative implementation.}
 
	\item We present a new {strong} seeded extractor construction, which we call $\Circulant$, {due to the input being used to construct a circulant matrix.}
    Beyond its simplicity, it is directly quantum-proof with the same error (and equivalently, the same output length)\footnote{The new construction is now a two-universal family of hash functions, allowing to apply well-known security proof techniques, as such, the results in \cite{renner2008security} prove it is quantum-proof with the same output length in the seeded case.}, making it the best choice against quantum adversaries. 
    It achieves the same entropy loss, quasi-linear {computation time} and error as Toeplitz-based ones \cite{krawczyk1994lfsr}, whilst requiring a seed that is {only a single bit longer than the weak input length} -- making it an excellent choice for tasks such as quantum key distribution. {This new extractor construction also has no alternative implementation.}
    % length as the weak input only
	
	\item We implement the $\Toeplitz$ extractor \cite{krawczyk1994lfsr} efficiently (in {$O(n_1\log(n_1))$}) and a $\Trevisan$ extractor, based on \cite{trevisan1999construction} and \cite{mauerer2012modular}, which tolerates the shortest seed lengths for sufficiently large input lengths but is less efficient {($O(m n_1\log^2 n_1)$ computation time, where $m$ is the output size of the extractor -- making it impractical in many cases)}\footnote{{We note that the seed length is $O(\log(n_1))$, but this does not always imply a short seed in practice. For instance, in our implementation with near-minimal entropy loss, the seed length exceeds the input length when $n_1 < 10^6$ and $m \approx n_1 / 2$. Furthermore, it is often useful to assume that the output length $m$ is proportional to the input length $n_1$, i.e., $m \propto n_1$. In this case, the computation time for the Trevisan extractor becomes $O(n_1^2 \log^2 n_1)$, making it inefficient. However, when $m$ is small, Trevisan's extractor can have a reasonable computation time, as seen in \cite{zhang2020experimental}, where $m = 512$.}} We also provide an implementation of the deterministic $\VonNeumann$ extractor in {$O(n_1)$}.
    
    \item {Contrary to all alternative implementations,} we implement the $\Circulant$, $\Dodis$ and $\Toeplitz$ extractors using the {number-theoretic} transform (NTT) {implemented in C++}, which gives a throughput of approximately $1$Mbit/s for input sizes up to $5 \times 10^{8}$ on a standard personal laptop (see \Cref{sec:performance})\@.
    Using the NTT instead of the fast Fourier transform (FFT) avoids numerical imprecision (due to floating point arithmetic) which may be unsuitable for some applications, e.g.\ in cryptography, especially with large input lengths\@.
    {Contrary to alternatives, our software is able to process input lengths of up to $2^{40} > 10^{12}$ bits, which should be sufficient even for device-independent protocols \cite{ekert1991quantum,acin2016certified,nadlinger2022experimental,shalm2021device,foreman2020practical,pironio2010random,bierhorst2018experimentally}}. 
    When using input lengths below $2^{29} \approx 5 \cdot 10^8$, the code uses a further optimised (smaller finite field for the) NTT to increase the throughput.
    
    \item We collate (and sometimes improve{, see for example \Cref{MarkovFor2Uni})} results from existing works into a single place, 
    providing techniques to get the most (near-)perfect randomness in different adversarial models and experimental settings. 
    For example, all our constructions can, as an option, extract under a weaker independence requirement on the weak seed (in the Markov model \cite{markov-extractors}) and we provide the associated parameters. 
    {Moreover, in the software we include \texttt{\.from\_params\(\)} utility functions which calculate the input and output lengths for a number of settings and a \texttt{suggest\_extractor} function, which assists a user on deciding which extractor to use for their application, based on the flow chart in \Cref{fig:ext-flow-chart}.}
\end{enumerate}

{To the best of our knowledge, existing works only achieve a small subset of the features mentioned above. There are no alternative implementations of the $\Dodis$ extractor or the $\Circulant$ extractor, which are new to this work.  
Reference~\cite{kharbanda20} is the only work we found that offers a library containing several software implementations of randomness extractors, including the $\VonNeumann$ and $\Toeplitz$ extractors, as well as one termed `universal hashing' and providing a link to the implementation of the $\Trevisan$ extractor \cite{mancusi2019}. However, it does not implement the $\Toeplitz$ extractor efficiently by exploiting the convolution theorem to obtain quasi-linear computation time, i.e., it does not use the FFT or NTT.} \\

Other software implementations of individual extractors exist. There are several publicly available implementations of the $\Toeplitz$ extractor, for example, in \cite{tanvirulz2018, BYUCamachoLab2022} (in Python) and \cite{rokzitko22} (in C++ and Verilog for FPGA), but none utilise the convolution theorem for quasi-linear computation time -- making them inefficient at larger input sizes. 
Implementations of the $\Trevisan$ extractor can be found in \cite{mauerer2014}, related to \cite{mauerer2012modular} and \cite{mancusi2019}, both written in C++. The implementation of \cite{mauerer2014} is particularly flexible, providing various instantiations of the extractor that can be useful to experts.
In \cite{aws2021}, the authors discuss and implement a modified $\Toeplitz$ extractor, based on \cite{hayashi-tsurumaru}, which is an extractor with parameters similar to $\Circulant$. This implementation uses the FFT, enabling input lengths up to $10^7$. However, by not using the NTT as we do, potential rounding errors may occur. A more detailed comparison is provided in \Cref{sec:toeplitz}.

\section{Overview of the extractor library}
\label{sec:randomness-extractor-implementations}
In this section, we present an overview of the \texttt{Cryptomite} extractors, an informal guide for selecting an appropriate extractor for an application, and performance benchmarking. 
$\Cryptomite$ is implemented in Python for usability and ease of installation, with performance-critical parts implemented in C++ called from Python.

\subsection{Extractors of \Cryptomite}
\label{sec:exts-of-cryptomite}
The $\Cryptomite$ library contains:
\begin{itemize}
	\item The (new) $\Circulant$ strong seeded extractor. It requires a prime with primitive root 2 seed length, $n_2=n_1+1$, and outputs $m = \lfloor k_1-2\log_2(\frac{1}{\epsilon})\rfloor$ bits against both classical and quantum adversaries (called classical- and quantum-proof respectively)\@. We present a two-source extension of this extractor, secure against quantum side information that is either product across the two sources (see \Cref{TwoExt_E2}) or follows the Markov model of side information \cite{markov-extractors} (see \Cref{TwoSource_E}), with output length $m=\lfloor k_1 + (k_2 - n_2) - 2\log_2(\frac{1}{\epsilon}) \rfloor$.
	\item The $\Dodis$ {strong} two-source extractor \cite{dodis2004improved}, with implementation based on \cite{foreman2020practical}\@. 
    It requires equal length inputs $n_1=n_2=n$, where $n$ is a prime with primitive root 2 and outputs $m = \lfloor k_1 + k_2-n-2\log_2(\frac{1}{\epsilon}) \rfloor$\@. 
    Its quantum-proof version in the Markov model outputs $m = \lfloor \frac{1}{5}(k_1 + k_2 - n + 8 \log_2 \epsilon + 9 - 4 \log_2(3)) \rfloor$\@.
	\item The $\Toeplitz$ strong seeded extractors based on \cite{krawczyk1994lfsr}\@. 
    It requires a seed length of  $n_2 = n_1 + m - 1$
	and outputs $m = \lfloor k_1-2\log_2(\frac{1}{\epsilon})\rfloor $, when classical-proof or quantum-proof with the same output length. We also present a two-source extension of this extractor, secure against quantum side information that is either product across the two sources or follows the Markov model of side information, with output length $m = \lfloor k_1 + (k_2-n_2)-2\log_2(\frac{1}{\epsilon}) \rfloor = \lfloor \frac{1}{2}(k_1 + k_2 - n_1 + 1 - 2\log_2(\frac{1}{\epsilon})) \rfloor$.
	\item The $\Trevisan$ {strong} seeded extractor \cite{trevisan1999construction}, with implementation based on \cite{mauerer2012modular}\@.
    It requires a seed length of $n_2 = O(\log(n_1))$ (see the exact statement of the seed length in \Cref{sec:trevisan}) and outputs $m = \lfloor k_1 - 6 - 4\log_2(\frac{m}{\epsilon}) \rfloor$ when classical-proof or quantum-proof with the same output length.
	As with $\Circulant$ and $\Toeplitz$, we offer two-source extensions, with the details deferred to \Cref{sec:trevisan}\@.
	\item The $\VonNeumann$ deterministic extractor \cite{von1963various}\@. 
     Although this extractor does not require a seed, the weak input must have more structure than min-entropy. 
     Additionally, this extractor incurs substantial entropy loss (see \Cref{sec:vn} for details)\@.
     More precisely, it requires that the weak input forms an \textit{exchangeable} sequence, e.g.\ a suitable weak input is one with IID bits. 
\end{itemize}

Our implementations of the $\Circulant$, $\Dodis$ and $\Toeplitz$ extractors all have $O(n_1\log(n_1))$ computation time and use the NTT\@.
For these extractors, our code can tolerate input lengths $n_1 \leq 2^{40} \approx 10^{12}$ bits -- with an improved throughput version for any input lengths under $2^{29}$ (for more details, see \Cref{sec:performance})\@.
{Our implementation of the $\VonNeumann$ extractor has a computation time of $O(n_1)$, while $\Trevisan$ has a computation time of $O\left(n_1 m \log^2\left(\frac{n_1 m}{\epsilon}\right)\right)$.}

\subsection{A simple user's guide}
\label{sec:practitioners-guide}
\begin{figure}[H] 
	\centering
	\includegraphics[width=0.8\textwidth]{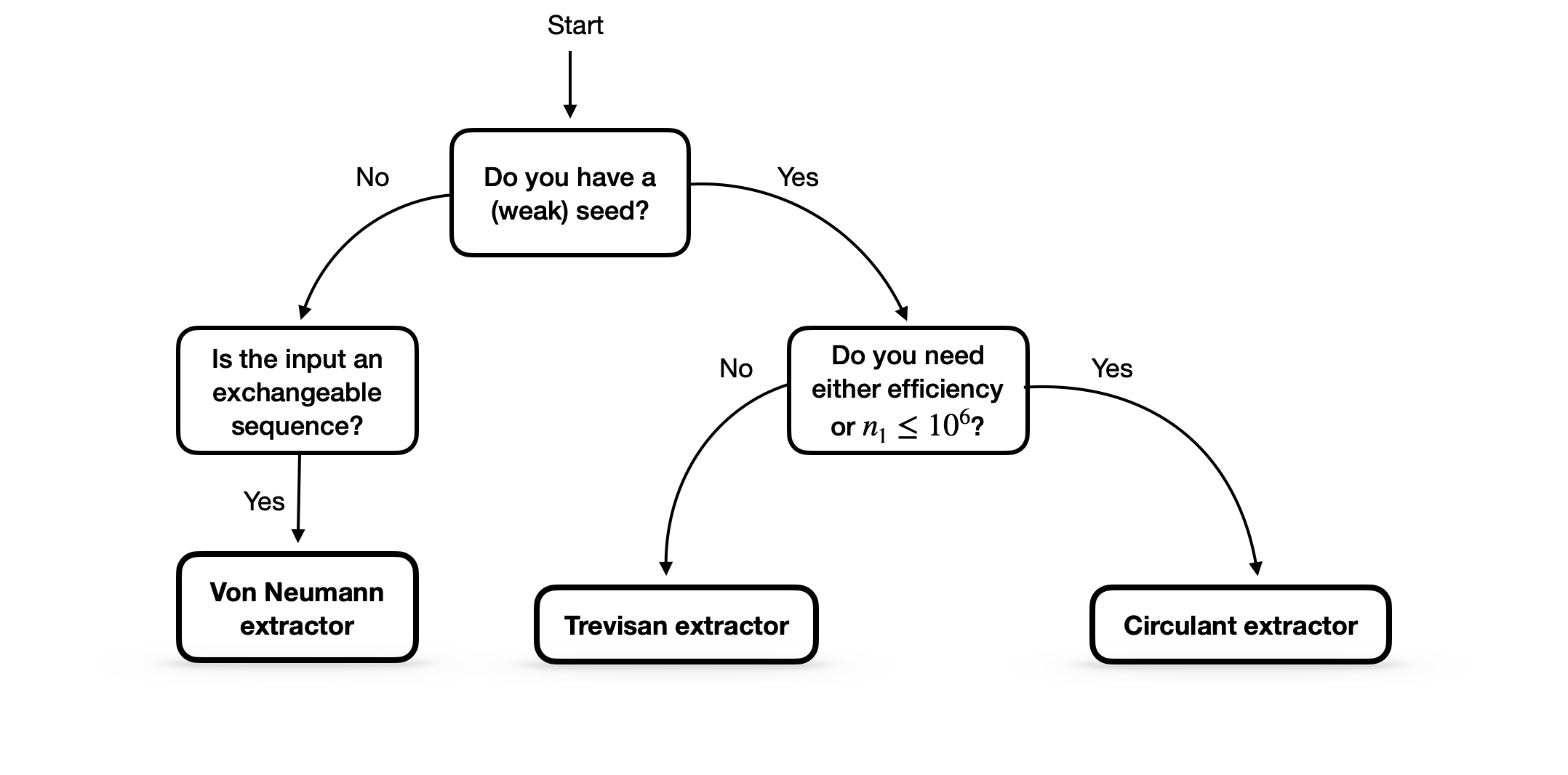}
  \caption{This flow chart shows the simplest way to get (approximately) the most output randomness from an input string of min-entropy $k_1$ and {length} $n_1$, whilst minimising the other resources (seed {length} and computation time)\@. Small improvements can be obtained by considering \Cref{fig:Table} below{, see for example a discussion on the seed lengths in \Cref{sec:comparison_seeded}}. In the left branch, the weak input bits need to form an exchangeable sequence, for which IID bits are a particular case.}\label{fig:ext-flow-chart}
\end{figure}

\begin{figure}[H] 
	\centering
	\includegraphics[width=1\textwidth]{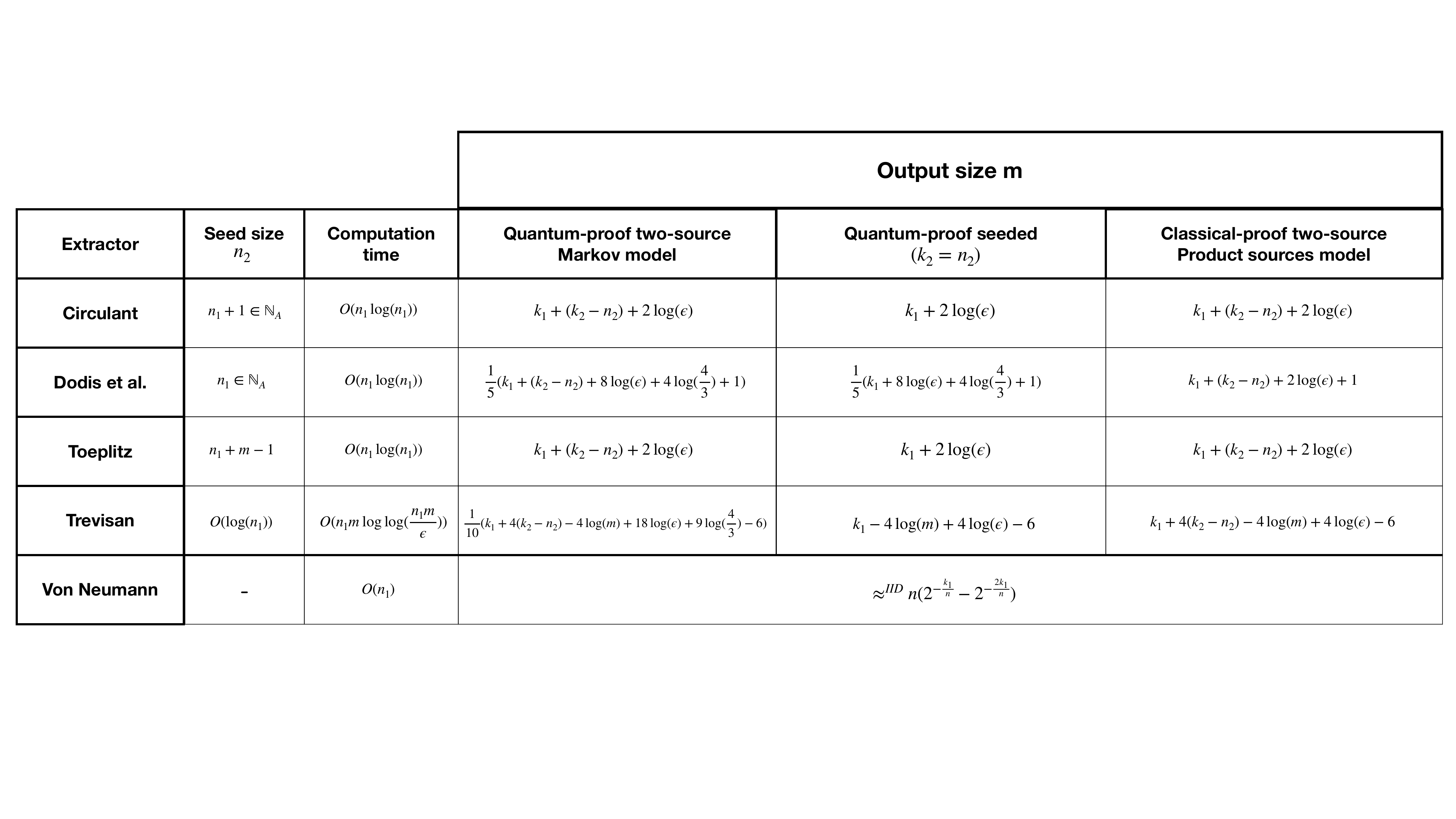}
	\caption{A summary of the parameters achievable by the different randomness extractors in $\Cryptomite$\@. For seeded extractors, one uses that $k_2=n_2$ and all logarithms are in base 2\@. $\mathbb{N}_A$ is the set of prime numbers with 2 as a primitive root. {The product source model refers to the two sources being independent (see \Cref{TwoExt_E2}) whilst the Markov model allows for the two sources to be correlated through a common cause (see \Cref{TwoSource_E}).} The parameters for classical {side information} in the Markov model are not included, but can be computed from \Cref{Markov1}\@. {We note that both Circulant and Toeplitz as two-source extractors are secure in the Markov model without a penalty (using \Cref{MarkovFor2Uni}), whilst Trevisan and Dodis require using the generic extension of \Cref{Markov2}. We have also added a discussion about the seed lengths of the different extractors in practice in \Cref{sec:comparison_seeded}.} Finally, the output length for $\VonNeumann$ is probabilistic and given in the case of IID bits as input (denoted $\approx^{IID}$)\@.
	}\label{fig:Table}
\end{figure}

To help using $\Cryptomite$, \Cref{fig:ext-flow-chart} presents a flowchart which assists a user on deciding which extractor to use for their application and \Cref{fig:Table} summarises the achievable parameters. 
The flowchart is somewhat informal and one can obtain small improvements by analysing each extractor individually (for example, finding a slightly longer output length or shorter seed {length})\@. 
The objective is to give a simple yet good choice for any application, in a clear and easy to follow diagram.
In this guide, the user begins with some extractor input of length $n_1$ and min-entropy $k_1$ (\Cref{def:min-ent})\@. {This flowchart is also implemented as a utility function $\texttt{suggest\_extractor}$ in $\Cryptomite$.}

\subsection{Performance} \label{sec:performance}
\begin{figure} [H]
  	\centering
  	\includegraphics[width=0.6\textwidth]{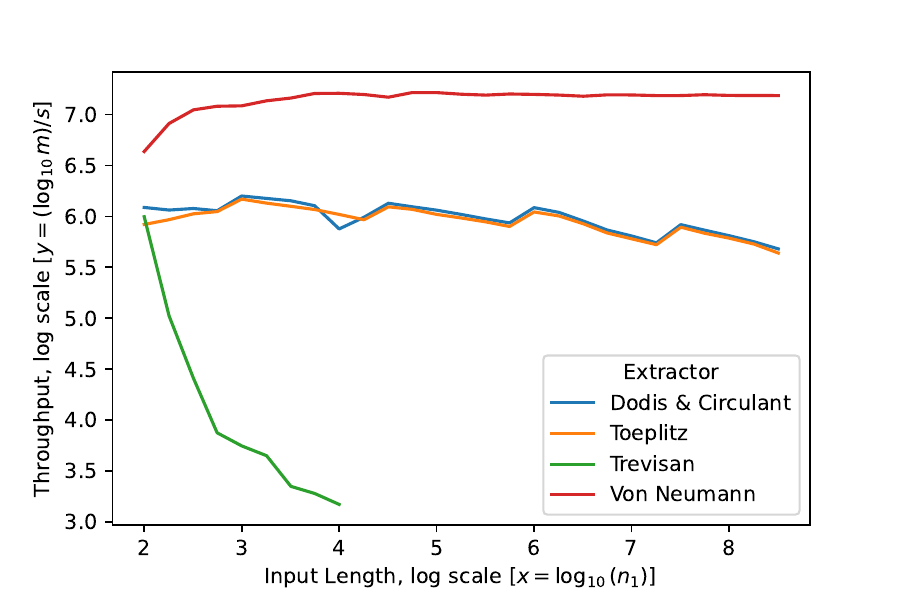}
  	\caption{A plot displaying the throughput of our extractor algorithms as a function of the input {length} ($n_1$) for each extractor of $\Cryptomite$, in logarithmic scale. The throughput is calculated by fixing the input min-entropy to $k_1 =\frac{n_1}{2}$ and calculating the maximum output length according to \Cref{fig:Table}. The necessary seed {length} for each extractor is given in \Cref{sec:library-in-detail}. \label{fig:performance}}
\end{figure}

To demonstrate the capabilities of \texttt{Cryptomite}, we perform some benchmarking of our constructions on a Apple M2 Max with 64GB RAM processor.
The \textit{throughput} (output bits per second) of the extractors of $\Cryptomite$ are shown in \Cref{fig:performance}\@.
It is calculated by averaging the run-time over 10 trials, for each input length, for input lengths up to $5 \times 10^8$.
Note that when the input length is below $2^{29} > 5 \times 10^8$ bits, the {convolution-based extractors} ($\Circulant$, $\Dodis$ and $\Toeplitz$) perform operations over a finite field defined by the prime $p = 3 \times 2^{30} + 1$. For input lengths above $2^{29}$ and up to $2^{40}$, the operations must be implemented in a larger finite field and we use the prime $p' = 9 \times 2^{42} + 1$ (which we call \texttt{bigNTT}). This change comes at the cost of approximately a 3-4$\times$ reduction in throughput.

Some observations are: 
\begin{itemize}
    \item The $\Circulant$, $\Dodis$\ and $\Toeplitz$\ extractors are able to output at speeds of up to $\approx 1$Mbit/s, even for large input lengths. The slight gradual decline in the larger input lengths is expected to be due to CPU characteristics such as cache hit rate and branch prediction. 
    \item The $\Trevisan$ extractor can generate output at speeds comparable to the $\Circulant$, $\Dodis$ and $\Toeplitz$ extractors only when the input or output {length} is (very) short. For input lengths greater than $n_1 \approx 10^4$, it could not generate a non-trivial throughput. 
    \item {The throughput of the $\VonNeumann$ extractor is lower for short input lengths due to a constant time overhead.}
  \end{itemize}

{We note that, for the $\Circulant$, $\Dodis$ and $\Toeplitz$ extractors, the run-time is independent of the output length ($m$) due to the NTT implementation step handling a square $n_1 \times n_1$ matrix before outputting the first $m$ bits. This means that the throughput in \Cref{fig:performance} could be doubled for those extractors if the input had min-entropy has been fixed to a value $k_1 \sim n_1$.}\\

{The code used to generate \Cref{fig:performance} can be found at \url{https://github.com/CQCL/cryptomite/blob/main/bench/bench.py}, enabling users to evaluate the performance of the extractors from $\Cryptomite$ on their specific device.}

\section{Theory of randomness extraction}
In this section, we introduce the notation and relevant definitions that we use for the remainder of the manuscript and provide some useful lemmas and theorems related to randomness extraction (which may be of independent interest)\@. 
The reader who is only interested in using our extractors can skip the preliminaries and go straight to the section for their extractor of choice and take the parameters directly, or use the repository documentation and parameter calculation modules that we make available.

\subsection{Preliminaries}
\label{sec:prelimaries}
We denote random variables using upper case, e.g.\ $X$, which take values $x$ in some finite alphabet with probability $\mathrm{Pr}(X=x) = p_{X}(x)$, and for some random variable $E$ taking values $e$, we write the conditional and joint probabilities as $\mathrm{Pr}(X=x | E=e) = p_{X|E=e}(x)$ and $\mathrm{Pr}(X=x,E=e) = p_{X,E}(x,e)$ respectively. {We say that two variables $X$ and $Y$ are independent if they are statistically independent, i.e. that $p_{X,Y}(x,y) = p_X(x)p_Y(y)$.}
We call the probability $p_{\mathrm{guess}}(X) = \max_{x} p_{X}(x)$ the (maximum) \textit{guessing probability} and the (maximum) conditional guessing probability $p_{\mathrm{guess}}(X|E) = \max_{x, e} p_{X|E=e}(x)$ (for more detailed discussions and explanations, see \cite{renner2006single})\@. 
In this work, we focus on the case that the extractor inputs are bit strings, e.g.\ $X \in \{0,1\}^n$ is a random variable of $n$ bits, and $X=x$ is its realisation.
Since the realisations $x$ are bit strings, we denote the individual bits with a subscript, e.g.\ $x = x_0, \ldots, x_{n-1}$ where $x_i \in \{0,1\}$ for all $i \in \{0, \ldots, n-1\}$\@. 
Let $[,]$ denote the concatenation of random variables, for example, if $X \in \{0,1\}^{n_1}$ and $Y \in \{0,1\}^{n_2}$, then $[X,Y] \in \{0,1\}^{n_1 + n_2}$ {is the concatenation of $X$ and $Y$}. 
Throughout the manuscript, all logarithms are taken in base 2 and all operations are performed modulo 2, unless explicitly stated otherwise. \\

{The concept of weak randomness, as we consider it, is captured by the (conditional) min-entropy. This can be interpreted as the minimum amount of random bits in a random variable, conditioned on the {{side information}} that a hypothetical adversary might possess.}
\begin{definition}[Conditional min-entropy] \label{def:min-ent}
  The (conditional) min-entropy $k$ of a {random variable} $X \in \{ 0,1\}^n$, denoted $\mathrm{H}_{\infty}(X|E)$, is defined as 
  \begin{align} k = \mathrm{H}_{\infty}(X|E) = -\log_2 \left(p_{\mathrm{guess}}(X|E) \right) \end{align}
  where $E$ is a random variable that contains any additional (or side) information that a hypothetical adversary may have to help predict $X$\@. {The total (conditional) min-entropy divided by its length is the \textit{min-entropy rate} $\alpha=\frac{k}{n}$}
\end{definition}

In order for randomness extractors to be useful in a cryptographic application, their output must satisfy a \textit{universally composable} \cite{canetti2001universally} security condition.
This condition is best understood by imagining a hypothetical game in which a computationally unbounded adversary is given both the \textit{real} output of the extractor and the output of an \textit{ideal} random source. 
The adversary wins the game if they can \textit{distinguish} between these two bit strings, i.e.\ guess which of the bit strings came from the extractor versus from the ideal random source. 
A trivial adversarial strategy is to guess at random, giving a success probability of $1/2$ and, for a secure protocol, we ask that a computationally unbounded adversary cannot distinguish the bit strings with success probability greater than $\frac{1}{2}+p_{\textrm{dist}}$, for $p_{\textrm{dist}} \ll 1$\@. 
In other words, the best strategy, even for a very powerful adversary, is not much better than the trivial one. 
This definition of security is formalised using distance measures as follows. 

\begin{definition}[Statistical distance] \label{def:stat-distance}
	The statistical distance $\Delta$ between two random variables $X$ and $Y$, with $X,Y \in \{0,1\}^n$, conditioned on information $E$ (a random variable of arbitrary dimension), is given by 
	\begin{align}
    \Delta\!\left(X, Y | E\right) 
        &= \frac{1}{2} \sum\limits_{x, e}  
		\lvert p_{X|E=e}(x) - p_{Y|E=e}(x) \rvert.
    \end{align}
\end{definition}
If $X$ and $Y$ satisfy $\Delta\!\left(X, Y | E\right) \leq \epsilon$ , we say that they are $\epsilon$-close (given $E$), which implies that they can be distinguished with advantage $p_{\textrm{dist}}\leq \frac{\epsilon}{2}$\@. 
This allows us to quantify how close a distribution is to uniform, which gives rise to our term (near-)perfect randomness, where near- is quantified by the distance bound.

\begin{definition}[Classical-proof (near-)perfect randomness] \label{def:classical-secure}
The random variable $Z \in \{0,1\}^m$ is (near-)perfectly random (given $E$), quantified by some value $\epsilon \in (0,1)$, if it is $\epsilon$-close to the uniform distribution $\mathrm{U}_{m}$ over $ \{0,1\}^m$, i.e.\
  \begin{align}
      \Delta\! \left(Z, \mathrm{U}_{m} | E\right) \leq \epsilon .
  \end{align}
\end{definition}
In our context, $\epsilon$ is called the extractor error. 
In practice, $\epsilon$ is a fixed security parameter, for example, enforcing that $\epsilon = 2^{-32}$\@. \\

We call the definitions that we have given so far \textit{classical-proof}, in the sense that the adversary's information is classical (a random variable $E$)\@. 
When the adversary has access to quantum states to store information, there are situations in which security cannot be guaranteed with \Cref{def:classical-secure} alone (see \cite{gavinsky2007exponential} for a concrete example and \cite{konig2011sampling} for a general discussion)\@.
Because of this, in what follows we generalise our definitions to the quantum case and say that these are secure against quantum adversaries, or \textit{quantum-proof}\@. \\

A random variable $X \in \{0,1\}^n$ with $p_X(x)$ can be written as a quantum state living in the Hilbert space $\X$ as $\rho_\X=\sum_{x\in \{ 0,1 \}^n} p_X(x) \ket{x}\!\! \bra{x}$, where the vectors $\{\ket{x}\}_{x}$ form an orthonormal basis. 
Considering quantum adversaries (with quantum side information), the classical variable $X$ can be considered part of a composite system potentially correlating it with the adversary in a Hilbert space $\X \otimes \E$, which can be written as the classical-quantum state
\begin{align}\label{def:rho XE}
	\rho_{\X\E} = 
	\sum_{x\in \{0,1\}^{n}} p_X(x)\ket{x}\!\! \bra{x}_\X \otimes \rho_{\E | x}.
\end{align} 
In \Cref{def:rho XE} the adversary's part of the system is the state $\rho_{\E | x}$, which is conditioned on the realisation $X=x$\@. 
In the quantum case, the conditional min-entropy (given $\E$) of the classical variable $X$ is $\mathrm{H}_{\infty}(X|\E)=-\log_2(p_{\mathrm{guess}}(X|\E))$, where $p_{guess}(X|\E) =\max_{\{M_x\}_{x}}\sum_{x} \Tr (M_x \rho_{\E | x})$ is the adversary's maximum guessing probability to guess $X$, given all measurement strategies $M_x$ on their system $\rho_{\E | x}$. 
For a general discussions on min-entropy in the presence of quantum adversaries, see \cite{konig2009operational}.\\

Similarly to the classical case, an extractor ideally outputs the classical-quantum state $\mathrm{u}_\Z \otimes \rho_{\E}$, where $\mathrm{u}_\Z$ is the maximally mixed state and independent of the state of the adversary
\begin{align} \label{def:ideal ZE}
  \mathrm{u}_\Z \otimes \rho_{\E} = 2^{-m} 
  \!\!\!\!\sum_{{z}\in \{0,1\}^{m}}\!\! 
  \ket{z}\!\! \bra{ z}_\Z \otimes \rho_{\E}. 
\end{align} 
{Analogously to the classical case, the quantum-proof security criterion is that the randomness extraction process produces a real state (\Cref{def:rho XE} with $Z$ as the classical subsystem) that is essentially indistinguishable from the ideal state in \Cref{def:ideal ZE}\@.} 
Intuitively, such an ideal state promises that the output $\mathbf{z}$ remains unpredictable for all adversarial strategies using their state $\rho_{\E | z}$\@.
This criterion for security against an adversary who has access to quantum {{side information}} is formalised in the following definition. 

\begin{definition}[Quantum-proof (near-)perfect randomness] \label{def:quantum-secure}
  The extractor output $Z \in \{0,1\}^m$ is quantum-proof $\epsilon$-perfectly random if
  \begin{align}\label{FORMARKOV}
     \Delta_Q (Z,\mathrm{U}_m|\E) = \frac{1}{2} \| \rho_{\Z \E} - \mathrm{u}_{\Z} \otimes \rho_{\E} \|_1  \leq \epsilon
  \end{align} where $\| \cdot \|_1$ denotes the trace norm, defined as $\| \sigma \|_1 = \mathrm{Tr}(\sqrt{ \sigma ^\dagger \sigma})$\@. 
  We use the subscript $Q$ and the conditioning on the quantum state $\E$ to mark the distinction with the classical version $\Delta(Z,\mathrm{U}_m|E)$\@.
\end{definition} For a discussion on why this is the correct measure against quantum adversaries, see \cite{renner2008security} `Section 2'\@.\\

We can now define what deterministic, seeded and two-source extractors are at the function level. 
For simplicity of notation, we omit the conditioning on the side information $E$ (or $\E$), for example, writing $\mathrm{H}_{\infty}(X)$ instead of $\mathrm{H}_{\infty}(X|E)$\@.

\begin{definition}[Deterministic randomness extractor]  
\label{def:det-extractor}
  A $(\epsilon, n, m)$-deterministic extractor is a function 
  \begin{align} 
    \mathsf{Ext_d}: \{ 0,1\}^n \rightarrow \{0,1\}^m 
  \end{align} 
  that maps a random bit string $X \in \{0,1\}^n$ with some specific properties\footnote{Such specific properties are, for example, that bits of $X$ are generated in an IID manner \cite{von1963various} or $X$ is a bit-fixing source \cite{kamp2007deterministic}\@. For a review of the properties sufficient for deterministic extraction, see \cite{shaltiel2011introduction}\@.}, to a new variable $Z=\mathsf{Ext_d}(X)$ that is (optionally quantum-proof) $\epsilon$-perfectly random, see \Cref{def:classical-secure}  (optionally \Cref{def:quantum-secure})\@. 
\end{definition}

\begin{definition}[Two-source randomness extractor]
\label{def:two-source-extractor}
Given two independent random variables $X \in \{0,1\}^{n_1}$ and $Y \in \{0,1\}^{n_2}$, with min-entropy $\mathrm{H}_{\infty}(X) \geq k_1$ and $\mathrm{H}_{\infty}(Y) \geq k_2$ respectively, a {$(n_1, \allowbreak k_1, \allowbreak n_2, \allowbreak k_2, \allowbreak m, \allowbreak \epsilon)$-two-source randomness extractor} is a function 
\begin{align} \mathsf{Ext_2}: \{ 0,1\}^{n_1} \times \{0,1\}^{n_2} \rightarrow \{0,1\}^m \end{align}
where $\mathsf{Ext_2}(X, Y)$ is $\epsilon$-perfectly random, see \Cref{def:classical-secure} (optionally \Cref{def:quantum-secure})\@.  
\end{definition} 
We discuss the role of {side information} for two-source extractors in \Cref{Useful}\@.
When discussing two-source extractors, we will refer to the input $X$ as the \textit{extractor input} and $Y$ as the \textit{weak seed}. 
In the case that $Y$ is already near-perfectly random, we will simply call it the \textit{seed}. 

\begin{definition}[Seeded randomness extractor] \label{def:seeded-extractor}
  A seeded extractor is a special case {$(n_1, k_1, n_2, k_2 = n_2, m, \epsilon)$}-two-source randomness extractor, where $k_2=n_2$.
\end{definition}

We have mentioned that our extractors are strong, and now we define this property 
\begin{definition}[Strong randomness extractor] \label{def:strong-extractor}
  A {$(n_1, k_1, n_2, k_2, m, \epsilon)$}-randomness extractor $\mathsf{Ext}: X \in \{0,1\}^{n_1} \times Y \in \{0,1\}^{n_2} \to \{0,1\}^m$ is strong in the input $Y$ if,   
  \begin{align} \Delta([\mathbf{\mathsf{Ext}}(X, Y), Y], [\mathrm{U}_{\mathrm{m}}, Y]) \leq \epsilon \end{align}
  or quantum-proof strong in the input $Y$ if 
    \begin{align} \Delta_Q ([\mathbf{\mathsf{Ext}}(X, Y), Y], [\mathrm{U}_{\mathrm{m}}, Y]) \leq \epsilon, \end{align}
  where $\mathrm{U}_{\mathrm{m}}$ denotes the uniform distribution over $m$ bits. 
  Informally, this can be understood as the property that the extractor output bits are near-perfect, even conditioned on the input $Y$.
\end{definition} 
By convention, we will always assume that the extractor is strong in the (weak) seed, although the role of the input sources can usually be exchanged for two-source extractors. 
All extractors presented in this work are strong with the same error and output length. Having this additional property leads to some useful consequences:
\begin{enumerate}
  \item $Y$ may become known to the adversary without compromising the security of the extractor output, i.e.\ the extractor output remains $\epsilon$-perfectly random.
  \item $Y$ can be re-used in multiple rounds of randomness extraction (with additive error $\epsilon$), under the assumption that each of the subsequent weak extractor inputs, say $X_2, X_3, \ldots $, are also independent of $Y$ (see \Cref{Thrm:seeds}).
  \item If $Y$ is near-perfectly random (the case of seeded extractors), it can be concatenated with the extractor output to increase the output length. If $Y$ only has min-entropy, it can be reused as the extractor input to another seeded extractor taking the output of the two-source extractor as seed. This {allows} to extract roughly all the min-entropy $\mathrm{H}_{\infty}(Y)$ (see \Cref{subsub:comp-exts}).
\end{enumerate}
Option 2 is compatible with options 1 and 3 (they can be achieved simultaneously), however option 3 is not compatible with 1 {because one should not use $Y$ in cryptographic applications (such as encryption for example) if it is known to an adversary}.

\subsection{Useful theorems and results}
\label{Useful}
Next, we reproduce and extend results from other works, which provides flexibility in the extractors' use.
First, we see how to account for {side information} with two-source extractors, where both sources might be correlated to the adversary's information. 
Second, we explain how to use extractors with a bound on the smooth min-entropy\footnote{The smooth min-entropy $\mathrm{H}_{\infty}^{\delta}(X)$ is the maximum min-entropy {of any random variable that is $\delta$ close (in terms of statistical distance) to the random variable $X$. For classical-quantum states, the closeness is in terms of trace distance.}}
only and how to account for near-perfect seeds (which can also be reused in multiple extraction rounds)\@. 
Third, we see how to generically extend a seeded extractor to a two-source one, as well as how to manipulate the input lengths to construct a family of extractors that can be optimised over. 
Finally, as mentioned above, we explain how one can concatenate two-source and seeded extractors in order to obtain more advantageous constructions.

\subsubsection{Two-source extractors in the product source model}
\label{TwoExt_E2}
Seeded randomness extractors are well studied in the presence of {side information} and quantum-proof constructions with minimal (or near minimal) entropy loss are known to exist \cite{renner2005universally,qproof-lhl,de2012trevisan,hayashi-tsurumaru}, where the weak input to the extractor is in a classical-quantum state, see \Cref{def:rho XE}.
Multi-source extractors, including two-source extractors require more care, as the adversary might now possess {side information} about both sources (for example potentially correlating the sources).\\

Hayashi and Tsurumaru \cite{hayashi-tsurumaru} showed that any strong classical-proof seeded extractor has a two-source extension, whereby the extractor can be used with a non-uniform seed at the cost of a larger error or shorter output length. 
Moreover, they give an improved extension when the extractor is based on two-universal families of hash functions\footnote{A family $\mathcal{F}$ of functions from $\mathcal{X} \in \{0,1\}^n$ to $\mathcal{Z} \in \{0,1\}^m$ is said to be \textit{two-universal} if $\Pr_{f}(f(x)=f(x'))\leq \frac{1}{m}$ for any $x\neq x' \in \mathcal{X}$ and $f$ chosen uniformly at random from $\mathcal{F}$ (which is the role of the seed). This definition can be extended beyond picking $f$ uniformly. See \cite{renner2008security} (section 5.4 and following) for more details and \cite{haastad1999pseudorandom} or the more recent \cite{barak2011leftover} that have shown that two-universal families of hash functions are good randomness extractors.}.

\begin{theorem}[Classical-proof two-source extension, Theorem 6 and 7 in \cite{hayashi-tsurumaru}]
  \label{lemma:2-weak-source}
  Any strong classical-proof (or quantum-proof) {$(n_1, k_1, n_2, k_2, m, \epsilon)$}-seeded extractor is a strong classical-proof (or quantum-proof) {$(n_1, k_1, n_2, k_2, m, 2^{n_2-k_2}\epsilon)$}-two-source extractor, strong in the (now) weak seed.
  If the extractor is constructed from a two-universal family of hash functions, it is a strong classical-proof (or quantum-proof) {$(n_1, k_1, n_2, k_2, m, 2^{(n_2-k_2)/2}\epsilon)$}-two-source extractor.
\end{theorem} 

The quantum-proof claim in \Cref{lemma:2-weak-source} does not directly cover quantum {side information} about the (now) weak seed, only on the weak input, and still demands that the weak seed and weak input are independent.
Using the proof techniques of \cite{berta2021robust} (`Proof of proposition 1') this can be directly generalised to the case that an adversary has quantum {side information} about both the input and the weak seed, but with the constraint that, for extractors based on two-universal hash families, this {side information} is in product form. 
More concretely, this is the case when the input and the now weak seed form a so-called classical-classical quantum-quantum state, whereby the combined two-universal hashing based extractor input (for the input and (weak) seed given by \Cref{def:rho XE}) takes the form
\begin{align}\label{ProductModelState}
    \rho_{XY\E_1\E_2} = \rho_{X\E_1} \otimes \rho_{Y\E_2}.
\end{align}
This model for quantum-proof extractors is the \textit{product source model}\@.

\begin{theorem}[Quantum-proof two-source extension in the product source model, \Cref{lemma:2-weak-source} with Proposition 1 in \cite{berta2021robust}]
  \label{lemma:2-weak-source-q}
  Any strong quantum-proof {$(n_1, k_1, n_2, k_2 = n_2, m, \epsilon)$}-seeded extractor based on two-universal hash families is a strong quantum-proof {$(n_1, k_1, n_2, k_2, m, 2^{(n_2-k_2)/2}\epsilon)$}-two-source extractor in the product source model.
\end{theorem}

\begin{proof}
    The proof of this follows directly from \cite{berta2021robust} `Proof of proposition 1' by noticing that $n-1$ can be generically replaced with $n_2$, the length of the (weak) seed and that their bound on $(*)$ (as appearing in the text) is found by bounding the collision entropy of the seeded version of the extractor, which holds for all extractors based off two-universal families of hash functions.
\end{proof}

\subsubsection{Two-source extractors in the Markov model}\label{TwoSource_E}
Two-source extractors can also be made secure in the so-called \textit{Markov model} \cite{markov-extractors}, which considers stronger adversaries than in the product source model.
Instead of requiring the standard independence relation between the input and weak seed, i.e.\ that the mutual information $\mathrm{I}(X:Y)=0$, one can use randomness extractors in the presence of {side information} that correlates the inputs by taking a penalty on the error or output length. The authors prove their results in the general cases of multi-source extractors, so we restate their theorems in the special case of two-source extractors.\\ 

In the case of classical {side information} $E$, with $\mathrm{I}(X:Y|E)=0$ instead of $\mathrm{I}(X:Y)=0$ and with conditional min-entropy's $\mathrm{H}_{\infty}(X|E)\geq k_1$ and $\mathrm{H}_{\infty}(Y|E)\geq k_2$; 
\begin{theorem}[Classical-proof in the Markov model, Theorem 1 in \cite{markov-extractors}]
	\label{Markov1}
	Any (strong) {$(n_1, k_1, n_2, k_2, m, \epsilon)$}-two-source extractor is a (strong) classical-proof {$(n_1, k_1-\log_2(1/\epsilon), n_2, k_2-\log_2(1/\epsilon), m, 3\epsilon)$}-two-source randomness extractor in the Markov model.
\end{theorem} 

In the quantum case (where the quantum {side information} of the adversary is now a quantum system $\E$) one requires $\mathrm{I}(X:Y|\E)=0$ instead of $\mathrm{I}(X:Y)=0$ and that both sources have conditional min-entropy, i.e.\ $\mathrm{H}_{\infty}(X|\E)\geq k_1$ and $\mathrm{H}_{\infty}(Y|\E)\geq k_2$\@. 

\begin{theorem}[Quantum-proof in the Markov model, Theorem 2 in \cite{markov-extractors}]
	\label{Markov2}
	Any (strong) {$(n_1, k_1, n_2, k_2, m, \epsilon)$}-two-source extractor is a (strong) quantum-proof {$(n_1, k_1-\log_2(1/\epsilon), n_2, k_2-\log_2(1/\epsilon), m, \sqrt{3\epsilon 2^{m-2}})$}-two-source randomness extractor in the Markov model.
\end{theorem}

{We now show that quantum-proof seeded extractors that are based on two-universal hash families can be transformed into two-source extractors in the Markov model with minimal penalty (only due to having a second weak source). This applies to both our Circulant construction and Toeplitz.}

\begin{corollary}\label{MarkovFor2Uni} {Any (strong) quantum-proof $(n_1, k_1, n_2, k_2 = n_2, m, \epsilon)$-seeded extractor based on two-universal hash families is also a (strong) quantum-proof $(n_1, k_1, n_2, k_2, m, 2^{(n_2-k_2)/2}\epsilon)$-two-source extractor in the Markov model.}
\end{corollary}

\begin{proof}
    {Via \Cref{lemma:2-weak-source-q}, the quantum-proof $(n_1, k_1, n_2, k_2 = n_2, m, \epsilon)$-seeded extractor becomes a quantum-proof $(n_1, k_1, n_2, k_2, m, 2^{(n_2 - k_2)/2} \epsilon)$ two-source extractor in the product source model, as it is based on two-universal hash families. According to \cite{hayden2004structure}, a Markov state $\rho_{XY\mathcal{E}}$ (for two sources $X$, $Y$, and the adversary's system $\mathcal{E}$, with the Markov conditions discussed in \Cref{Markov2}) can be decomposed into product states (\Cref{ProductModelState}) as
    \begin{align}
        \rho_{XY\mathcal{E}} = \bigoplus_t p(t) \rho^t_{X\mathcal{E}_X^t} \otimes \rho^t_{Y\mathcal{E}_Y^t}
    \end{align} where $t$ runs over a finite alphabet with probability distribution $p(t)$, and $\mathcal{H}_\mathcal{E} = \bigoplus_t \mathcal{H}_{\mathcal{E}_X^t} \otimes \mathcal{H}_{\mathcal{E}_Y^t}$ is the Hilbert space of $\mathcal{E}$ (the quantum system held by the adversary). Using this decomposition, we can rewrite \Cref{FORMARKOV} as
    \begin{align}\label{FreeMarkov}
        \left\| \rho_{\text{Ext}(X,Y)\mathcal{E}} - u_\mathcal{Z} \otimes \rho_\mathcal{E} \right\|_1
        &=\left\| \text{Ext} \otimes \mathbb{I}_\mathcal{E} \left( \rho_{XY\mathcal{E}} \right) - u_\mathcal{Z} \otimes \rho_{\mathcal{E}} \right\|_1 \\
        &=\left\| \text{Ext} \otimes \mathbb{I}_\mathcal{E} \left( \bigoplus_t p(t) \rho^t_{X\mathcal{E}_X^t} \otimes \rho^t_{Y\mathcal{E}_Y^t} \right) - u_\mathcal{Z} \otimes \left(\bigoplus_{t}\rho^t_{\mathcal{E}_X^t} \otimes \rho^t_{\mathcal{E}_Y^t}\right) \right\|_1 \\
        &= \sum_t p(t) \left\| \text{Ext} \otimes \mathbb{I}_\mathcal{E} \left( \rho^t_{X\mathcal{E}_X^t} \otimes \rho^t_{Y\mathcal{E}_Y^t} \right) - u_\mathcal{Z} \otimes \rho^t_{\mathcal{E}_X^t} \otimes \rho^t_{\mathcal{E}_Y^t} \right\|_1 \\
        &\leq \sum_{t} p(t) 2^{(n_2 - k_2)/2} \epsilon = 2^{(n_2 - k_2)/2} \epsilon
    \end{align}
    where $\text{Ext}(X,Y)$ denotes the application of the extractor on $X,Y$, which is also understood as the CPTP map $\text{Ext} \otimes \mathbb{I}_\mathcal{E}$ on $\rho_{XY\mathcal{E}}$, and $u_\mathcal{Z}$ is the maximally mixed state over the output $Z$ of the extractor. In the last inequality, we have used the fact that the extractor is already quantum-proof in the product source model, meaning that, $\left\| \text{Ext} \otimes \mathbb{I}_\mathcal{E} \left( \rho^t_{X\mathcal{E}_X^t} \otimes \rho^t_{Y\mathcal{E}_Y^t} \right) - u_\mathcal{Z} \otimes \rho^t_{\mathcal{E}_X^t} \otimes \rho^t_{\mathcal{E}_Y^t} \right\|_1 \leq 2^{(n_2 - k_2)/2} \epsilon$ for all $t$, noting that the ccq-state $\rho^t_{X\mathcal{E}_X^t} \otimes \rho^t_{Y\mathcal{E}_Y^t}$ satisfies $\mathrm{H}_{\infty}(X|\mathcal{E}) \geq k_1$ and $\mathrm{H}_{\infty}(Y|\mathcal{E})$ for any $t$, given $\mathrm{H}_{\infty}(X|\mathcal{E}) \geq k_1$ and $\mathrm{H}_{\infty}(Y|\mathcal{E})$ for $\rho_{XY\mathcal{E}}$.}
\end{proof}

{In both Theorems~\ref{Markov1}, \ref{Markov2} and \Cref{MarkovFor2Uni}, one has the option to lift an extractor to the Markov model whilst maintaining whether it is strong or not (a strong extractor will be strong in the Markov model, a weak one will remain weak). We have indicated this option by putting strong in parenthesis in the statements.\\}

The Markov model is particularly relevant in protocols for randomness amplification \cite{kessler2020device, foreman2020practical} and in quantum key distribution (QKD)\footnote{In quantum key distribution (QKD), if the adversary has some predictive power over the seed that is used in the privacy amplification step, one could also imagine that its actions on the quantum channel correlates the raw key material and the (now weak) seed. Such situations, in which the different sources become correlated through a third variable or system, can be accounted for by working in the Markov model.}. 
Other works have considered different forms of quantum {side information}, for example, where the adversary is allowed specific operations to build the {side information} $\E$ \cite{kasher2010two} or when there is a bound on the dimension of the adversary's {side information} \cite{ta2009short}\@. 
However, most of these other forms of {side information} can be viewed as a particular case of the Markov one, making it a useful generic extension to use -- see the discussion in \cite{markov-extractors} (Section 1.2: `Related work')\@.
In the classical case, \cite{ball2022randomness} explores what happens to the security of two-source extractors when the extractor input and the weak seed can be correlated in some specific models, for example, when they have bounded mutual information $I(X:Y) \leq t$, for some constant $t$\@.

\subsubsection{Working with near-perfect seeds and smooth min-entropy bounds}

Seeded randomness extraction can be performed using seeds that are near-perfectly random only (instead of perfectly random), making it more practical. 
Strong seeded extractors allow the seed to be reused multiple times, and so we give a theorem that accounts for this seed reuse and we discuss this more in \Cref{subsub:comp-exts}\@.

\begin{theorem}[Extractors with {near-perfect} seeds, Appendix A in \cite{frauchiger4547true}] \label{Thrm:seeds}
	Given an $\epsilon_s$-perfect seed $Y$, which is used to extract $t$ times from a weak input\footnote{Called a block min-entropy condition. The theorem can be generalised to the case that the min-entropy of each block differs, i.e.\ $\mathrm{H}_{\infty}(X_i|X_{i-1},...,X_{1}) \geq k_i$\@.} with $X_i$ for $i=1,...,t$ such that $\mathrm{H}_{\infty}(X_i|X_{i-1},...,X_{1}) \geq k$ for all $i$ using a strong $(n_1, k_1 = k,n_2, k_2 =n_2, m, \epsilon)$-seeded extractor $\mathsf{Ext_s}(X,Y)$ each time, the concatenated output $[\mathsf{Ext_s}(X_1,Y),...,\mathsf{Ext_s}(X_t,Y)]$ is $\epsilon_t$-perfect with
	\begin{align}
		\epsilon_t \leq t \cdot \epsilon + \epsilon_s.
	\end{align}
\end{theorem}

Similarly, in practice, often only a bound on the smooth min-entropy can be obtained, for example when using the (generalised) entropy accumulation theorem \cite{dupuis2020entropy,metger2022generalised,dupuis2019entropy,arnon2018practical} or probability estimation factors \cite{zhang2018certifying,knill2020generation,zhang2020efficient}. 
We give two theorems that allow for seeded and two-source extractors to function with inputs that have a promise on their smooth min-entropy.
\begin{theorem}[Seeded extraction with smooth min-entropy, e.g. Corollary 7.8 in \cite{tomamichel2015quantum}]
	Given a (strong) quantum-proof $(n_1, k_1=k,n_2,k_2 = n_2, m, \epsilon)$-seeded extractor, one can use a weak input $X$ with smooth min-entropy $\mathrm{H}_{\infty}^{\delta}(X|\E) \geq k$ at the cost of an additive error $\epsilon+\delta$, {i.e.\ giving a $(n_1, k_1 =k,n_2, k_2 = n_2, m, \epsilon+\delta)$-seeded extractor.}
\end{theorem}

\begin{theorem}[Two-source extraction with smooth min-entropy, Lemma 17 in \cite{arnon2018practical}]
	Given a (strong) quantum-proof $(n_1, k_1-\log_2(1/\epsilon_1)-1, n_2, k_2-\log_2(1/\epsilon_2)-1,m, \epsilon)$-two-source extractor in the Markov model (see \Cref{TwoSource_E}), one can use weak inputs $X$ and $Y$ satisfying $\mathrm{I}(X:Y|\E)=0$ with smooth min-entropy's $\mathrm{H}_{\infty}^{\delta_1}(X|\E) \geq k'_1$ and $\mathrm{H}_{\infty}^{\delta_2}(Y|\E) \geq k'_2$ giving a
	\begin{align}
		\left(n_1, k'_1-\log_2 \left(\frac{1}{\epsilon_1} \right),n_2, k'_2-\log_2 \left(\frac{1}{\epsilon_2}\right),m , 2\epsilon+6(\delta_1+\delta_2)+2(\epsilon_1+\epsilon_2)\right)\mathrm{-extractor}
	\end{align}
\end{theorem}
This result also applies to classical-proof extractors and, because of the generality of the Markov model, to many other forms of (quantum) {side information} \cite{markov-extractors}.

\subsubsection{Extending extractors to more settings}
As some of our extractors require specific input lengths (for example, that the input length must be prime with primitive root 2), we give some results that give flexibility to increase or decrease the length of the extractor input, which later allows us to optimise the output length.
First, we show that any two-source extractor can be used as a seeded extractor with a short seed, by concatenating the seed with fixed bits.

\begin{theorem}[Short seeded extractors from two-source {extractors}] \label{thm:short-seed-ext}
    {Any $(n_1, k_1, n_2, k_2, m, \epsilon)$-two-source randomness extractor $\mathsf{Ext_2}$
    is a $(n_1, k_1, n_2'=k_2, k_2, m, \epsilon)$-seeded randomness extractor $\mathsf{Ext_s}$ for any $k_2 \leq n_2$, where $\mathsf{Ext_s}$ is constructed by first concatenating the seed (of length $k_2$) with a fixed bit string of length $n_2 - k_2$ and then applying $\mathsf{Ext_2}$. Specifically, given a seed $Y \in \{0,1\}^{k_2}$, the seeded extractor is constructed from the two-source one as $\mathsf{Ext_s}(X,Y) = \mathsf{Ext_2}(X,[Y,c])$, where $[Y,c]$ denotes the concatenation of the seed $Y$ with any fixed string $c \in \{0,1\}^{n_2- k_2}$ (e.g.\ $c=\{0\}^{n_2-k_2}$).}
\end{theorem}

\begin{proof}
    {Let $Y \in \{0,1\}^{k_2}$ be a seed (i.e. it has full min-entropy $\mathrm{H}_{\infty}(Y) = k_2$) and $c \in \{0,1\}^{n_2 - k_2}$ be a fixed string, e.g. $\{0\}^{n_2 - k_2}$, with $\mathrm{H}_{\infty}(c) = 0$.
    Define $Y'$ as the concatenation of $Y$ and $c$, $Y' = [Y, c]$, which has min-entropy $\mathrm{H}_{\infty}(Y') = \mathrm{H}_{\infty}([Y, c]) = \mathrm{H}_{\infty}(Y) = k_2$.
    Since $Y'$ is a random variable with min-entropy $k_2$ and length $n_2$, it can be used as the second input to any $(n_1, k_1, n_2, k_2, m, \epsilon)$-two-source extractor $\mathsf{Ext_2}(X, Y')$. 
    Therefore, $\mathsf{Ext_2}(X, Y')$ can be understood as a seeded extractor $\mathsf{Ext_s}(X, Y)$, constructed by first concatenating the seed $Y$ with a fixed string of bits then applying $\mathsf{Ext_2}$. In other words, $\mathsf{Ext_s}(X, Y) = \mathsf{Ext_2}(X, [Y,c]) = \mathsf{Ext_2}(X, Y')$.}
\end{proof}

{This theorem is useful when a seed is too short to be used in a seeded extractor for a given weak input. For example, consider a seed $Y$ of length $n_2$ and a weak input $X \in \{0,1\}^{n_1}$ with $n_1 > n_2$. If $n_1$ is sufficiently large compared to $n_2$, the seed length $n_2$ may be insufficient to perform extraction using existing seeded extractor constructions. However, by concatenating $Y$ with a zero string of length $c$, one can create a weak seed $Y'$ of length $n_2 + c$ with the same min-entropy $k_2 = n_2$. This weak seed is now a valid second input for a two-source extractor, which may give more freedom and allow extraction to become possible.}

We now show how to shorten the weak input {length}, if desired.

\begin{lemma}\label{lemma:shorten-entropy}
  {For any integer $c \leq n$, a random variable $X \in \{0,1\}^{n}$ with min-entropy $\mathrm{H}_{\infty}(X) = k$ can be shortened to a new random variable $X' \in \{0,1\}^{n - c}$ with min-entropy $\mathrm{H}_{\infty}(X') \geq k - c$, by removing the last $c$ bits of $X$.}
  %_{0:n-c-1}
\end{lemma}

\begin{proof}
 {Let $X' \in \{0,1\}^{n_1-c}$, $\bar{X}' \in \{0,1\}^c$ and $X = [X', \bar{X}'] \in \{0,1\}^{n_1}$, where $[,]$ denotes concatenation. By definition, $\mathrm{H}_{\infty}(X) = k$ and we want to prove a lower bound on $\mathrm{H}_{\infty}(X')$.
 Using that $p_{X'}(x')$ is a marginal distribution of $p_{[X', \bar{X}']}([x', \bar{x}'])$, we have that \begin{align} \label{eq:stop}
  \mathrm{H}_{\infty}(X') &= -\log_2(\max\limits_{x'} p_{X'}(x') )
  =-\log_2\left(\max\limits_{x'} \sum\limits_{\bar{x}' \in \{0,1\}^{c}} p_{[X', \bar{X}']}([x', \bar{x}']) \right). 
  \end{align}
  Continuing from \Cref{eq:stop} by passing the maximum inside the sum and then maximise over both variables (instead of one), we get
  \begin{align}
  -\log_2\left(\max\limits_{x'} \sum\limits_{\bar{x}' \in \{0,1\}^{c}} p_{[X', \bar{X}']}([x', \bar{x}']) \right) \geq -\log_2 \left( \sum\limits_{\bar{x}' \in \{0,1\}^{c}} \max\limits_{x'} p_{[X', \bar{X}']}([x', \bar{x}']) \right) \\
  \geq -\log_2 \left( \sum\limits_{\bar{x}' \in \{0,1\}^{c}} \max\limits_{[x', \bar{x}']} p_{[X', \bar{X}']}([x', \bar{x}']) \right) = 
   -\log_2 \left( \sum\limits_{\bar{x}' \in \{0,1\}^{c}} \max\limits_{x} p_{X}(x) \right) \\
  \geq -\log_2 \left( \sum\limits_{\bar{x}' \in \{0,1\}^{c}} 2^{-k} \right) 
  = -\log_2(2^{-k} 2^{c}) = k-c \end{align}
  where we have used that $\max\limits_{x} p_{X}(x) \leq 2^{-k}$, since $\mathrm{H}_{\infty}(X) = k$ by definition.}
\end{proof}

\subsubsection{Composing extractors} \label{subsub:comp-exts}
Extractors can be composed together to form new extractors with improved parameters. 
One can append a strong seeded extractor to a strong two-source extractor, taking the two-source extractor output as a seed and the two-source weak seed as the extractor input.
Doing so, one can construct a $(n_1, k_1, n_2, k_2, m + \max\{k_1,k_2\}, \epsilon_2 + \epsilon_s)$-two-source extractor, which significantly increases the output length. 
For more details, see Lemma 38 in \cite{markov-extractors} and the discussion in Section 4.6 of \cite{foreman2020practical}\@.

\section{Library in detail}
\label{sec:library-in-detail}
In this section, we concretely define the extractors in $\Cryptomite$, providing implementation details and parameter calculations.
\subsection{Our new $\Circulant$ construction}
We now expose the details of our new extractor construction, $\Circulant$\@. 

\begin{definition} \label{ext-circ}
	Let $x = x_0, \ldots, x_{n-2} \in \{0,1\}^{n-1}$ and $y = y_0, \ldots, y_{n-1} \in \{0,1\}^n$ where $n$ is prime with primitive root 2.\footnote{A list of all primes with 2 as a primitive root up to $10^6$ can be found at \href{https://github.com/CQCL/cryptomite/blob/main/na_set.txt}{https://github.com/CQCL/cryptomite/blob/main/na\_set.txt}\@. We also provide a function which allows the user to compute the closest prime with 2 as primitive root, as well as individually the closest above and below the input length, see the utility functions \href{https://github.com/CQCL/cryptomite/blob/main/cryptomite/utils.py}{https://github.com/CQCL/cryptomite/blob/main/cryptomite/utils.py}\@.}
	The function $\Circulant(x,y):\{0,1\}^{n-1} \times \{0,1\}^n \rightarrow \{0,1\}^m$ is implemented by: 
	\begin{enumerate}
		\item Set $x' = [x, 0] \in \{0,1\}^n$, where $[,]$ denotes concatenation (i.e., $x'_{n-1}=0$)\@.
		\item Then,
		\begin{align}
			\Circulant(x, y) = \left(\mathrm{circ}(x') y \right)_{0:m-1},
		\end{align} where the matrix-vector multiplication $\mathrm{circ}(x') y$ is taken mod 2 (in each component of the resulting vector) and the subscript ${0:m-1}$ denotes the first $m$ elements of the vector. The term $\mathrm{circ}(x')$ is the $n \times n$ circulant matrix generated by $x' = x'_0, \ldots x'_{n-1}$,
		\begin{align} \label{eq:circ-matrix}
			\mathrm{circ}(x') = 
			\begin{bmatrix}
				x'_0 & x'_1 & x'_2 & \ldots & x'_{n-2} & x'_{n-1} \\
				x'_{n-1} & x'_0 & x'_1 & x'_2 & \ldots & x'_{n-2} \\
				& \ddots & \ddots & \ddots & \ddots & & \\
				x'_1 & x'_2 & x'_3 & \ldots & x'_{n-1} & x'_0
			\end{bmatrix} , 
		\end{align} 
	\end{enumerate}
\end{definition}

Using the result in the Appendix of \cite{foreman2020practical}, the function in \Cref{ext-circ} can be implemented in $O(n\log n)$ computation time. 
\begin{theorem}\label{circSEED}
	The function $\Circulant(X, Y):\{0,1\}^{n-1} \times \{0,1\}^n \rightarrow \{0,1\}^m$ in \Cref{ext-circ} is a strong in $Y$, classical-proof and quantum-proof {$(n_1 = n-1, k_1, n_2 = n, k_2 = n, m, \epsilon)$-seeded extractor} for prime with primitive root 2 $n$, with output length 
	\begin{align}
		m \leq k_1 + 2\log_2(\epsilon).
	\end{align}
\end{theorem}
\noindent The proof can be found in \Cref{CirculantConstruction}\@.\\

Using \Cref{lemma:2-weak-source} to build a two-source extension that is (classical-proof and) quantum-proof in the product source model, our $\Circulant$ extractor has an output length 
	\begin{align}
		m \leq k_1 + k_2 - n + 2\log_2(\epsilon)
	\end{align} and, by \Cref{MarkovFor2Uni}, quantum-proof in the Markov model with the same output length.

\subsection{Dodis et al.}
\label{sec:dodis}
The exact $\Dodis$ extractor \cite{dodis2004improved} $\Dodis(x,y): \{0, 1\}^n \times \{0, 1\}^n \to \{0, 1\}^m$ construction is given in \Cref{subsec:dodis-imp}, whilst we give its idea here. 
The extractor input $x = x_0, x_1, \ldots, x_{n - 1} \in \{0,1\}^{n}$, and the (weak) seed $y = y_0, y_1, \ldots, y_{n - 1} \in \{0,1\}^{n}$ must both be of equal length $n$\@.
The extractor construction uses a set of $m$ $n \times n$ matrices, which we label $A_0,...,A_{m-1}$, with entries in $\{0,1\}$ (i.e.\ bits), which must be chosen such that, for any subset $B \subseteq A_0,...,A_{m-1}$, the sum of the matrices in the subset $B$ has rank {at least} $n-r$ for some small constant $r$ (which acts as a penalty to the output length)\@.
The extractor output of $m$ bits is then
{\begin{align}\label{DodisOUT}
  \Dodis(x,y) = [ (A_0 x) \cdot y, (A_1 x) \cdot y, ... , (A_{m-1} x) \cdot y] 
\end{align}}
where $\cdot$ is the inner product modulo 2\@. 
We give such a set of matrices, based on ``Cyclic Shift matrices'' from \cite{dodis2004improved} Section 3.2, and a version of \Cref{DodisOUT} that can be implemented in $O(n \log(n))$ computation time, based on \cite{foreman2020practical} in \Cref{subsec:dodis-imp}\@.
This construction requires that $x \neq \{0\}^n$ or $\{1\}^n$ and that $n$ is a prime with 2 as a primitive root. 
For the input lengths considered in this work (i.e.\ up to $10^{12}$), this set is sufficiently dense in the natural numbers for all practical purposes. 
We also note that fixing the input length is a pre-computation that can be done very efficiently. \\

We note that the particular choice of Cyclic Shift matrices $\{A_i\}_i$, together with the padding of step 1 in \Cref{ext-circ}, gives the $\Circulant$ construction -- which we prove to be a two-universal family of hash functions and in turn use known proof techniques and results, for example, giving that $\Circulant$ is quantum-proof without a penalty \cite{renner2008security}\@. 
\begin{theorem}[Classical-proof, from \cite{dodis2004improved}]
	Our implementation of the $\Dodis$ extractor is a strong classical-proof $(n_1 = n, k_1, n_2 = n, k_2 = n, m, \epsilon)$-seeded randomness extractor with
	\begin{align}
		m \leq k_1 + 1 + 2 \log_2 \epsilon,
	\end{align}
	and a classical-proof $(n_1 = n, k_1, n_2 = n, k_2, m, \epsilon)$-two-source randomness extractor, strong in the {weak seed (with min-entropy $k_2$)}, with
	\begin{align}
		m \leq k_1 + k_2 - n + 1 + 2 \log_2 \epsilon.
	\end{align}
\end{theorem}
We note that the $\Dodis$ extractor was shown to be a two-source extractor directly \cite{dodis2004improved}, i.e.\ without using a generic extension, and therefore outputs one more bit than other constructions in this setting.
Contrary to the $\Circulant$ construction, $\Dodis$ is not known to be quantum-proof directly (i.e.\ without a penalty) even as a seeded extractor.
Therefore, as explained in \Cref{TwoSource_E}, we make it quantum-proof in the Markov model \cite{markov-extractors} by taking a penalty on the output length.
 
\begin{theorem}[Quantum-proof in the Markov model, Proposition 5 of Appendix B in \cite{foreman2020practical}]
	The $\Dodis$ extractor becomes a strong quantum-proof (in the Markov model)  $(n_1 = n, k_1, n_2 = n, k_2=n, m, \epsilon)$-seeded randomness extractor, with
	\begin{align} 
		m \leq \frac{1}{5} \left(k_1 + 8 \log_2 \epsilon + 4 \log_2 \left(\frac{4}{3}\right) +1 \right), 
	\end{align}
	and a strong quantum-proof (in the Markov model) $(n_1 = n, k_1, n_2 = n, k_2, m, \epsilon)$-two-source randomness extractor, where
	\begin{align} 
		m \leq \frac{1}{5} \left( k_1 + k_2 - n + 8 \log_2 \epsilon +  4 \log_2 \left(\frac{4}{3}\right) +1 \right) .
	\end{align}
\end{theorem}

\subsection{Toeplitz}
\label{sec:toeplitz}
The $\Toeplitz$ extractor \cite{krawczyk1994lfsr} $\Toeplitz(x,y): \{0, 1\}^{n} \times \{0, 1\}^{n+m-1} \to \{0, 1\}^m$ is constructed using Toeplitz matrices. 
Given $x = x_0, x_1, \ldots x_{n-1} \in \{0,1\}^n$ and $y = y_0, y_1, \ldots y_{n+m-2} \in \{0,1\}^{n+m-1}$, the output of the extractor is computed as the matrix-vector multiplication
\begin{align} \label{eq:toeplitz-ext}
\Toeplitz(x,y) =\mathrm{toep}(y) x
\end{align}
where {\begin{align} \label{eq:toep}
    \mathrm{toep}(y) = \begin{bmatrix}
    y_0 & y_{n + m - 2} & \ldots  & \ldots& y_{m} \\
    y_{1} & y_0 & \ddots  & \ddots & \vdots \\
    \vdots & \vdots &\ddots & \ddots & \vdots \\
    y_{m-1} & y_{m - 2} & \ldots & \ldots &  . 
    \end{bmatrix} 
\end{align}} is the $n \times m$ Toeplitz matrix generated from $y$\@.
The full implementation details, in computation time $O(n \log(n))$, is given in \Cref{subsec:toeplitz-imp}\@. 
As it was shown to be directly quantum-proof \cite{impagliazzo1989pseudo, renner2008security}, we state the claims as a single theorem.

\begin{theorem}
	The $\Toeplitz$ extractor is a strong classical- and quantum-proof $(n_1 = n, k_1, n_2 = n + m -1, k_2= n + m -1, m, \epsilon)$-seeded randomness extractor, where
	\begin{align}
		m \leq k_1 + 2 \log_2(\epsilon), 
	\end{align} 
\end{theorem}

Using the two weak sources extension (\Cref{lemma:2-weak-source}), a classical- and quantum-proof $(n_1 = n, k_1, n_2 = n + m -1, k_2, m, \epsilon)$-two-source randomness extractor strong in the {weak seed (with min-entropy $k_2$)}, where
\begin{align} \label{params:toeplitz-2}
    m \leq \frac{1}{2} (k_1 + k_2 - n + 1 + 2\log_2(\epsilon)),
\end{align} in the product source model and {also in the Markov model (by \Cref{MarkovFor2Uni}).} Note that $k_2$ may be bigger than $n$, due to the fact that the length of the (weak) seed is $n + m -1$\@. \\

\noindent \textit{Modified Toeplitz Extractor -- }
To reduce the seed length, one can use modified Toeplitz matrices (Subsection B-B in \cite{hayashi-tsurumaru}), which allow a reduction from $n+m-1$ to $n-1$\@. 
The authors also prove the security of the two-source extension against quantum {side information} on the weak input only (which we reproduced as \Cref{lemma:2-weak-source}) and in \cite{berta2021robust} the security is extended against product quantum side information (i.e.\ {side information} on both inputs)\@. 
One could further extend it to allow for more general forms of quantum side information (in the Markov model) using \Cref{Markov2}\@.

\subsection{Trevisan}
\label{sec:trevisan}
The $\Trevisan$ extractor \cite{trevisan1999construction} $\Trevisan(x,y): \{0, 1\}^n \times \{0, 1\}^{d} \to \{0, 1\}^m$ is important because of its asymptotic logarithmic seed length $d=O(\log(n))$\@. 
At a high level, it is built from two components: (1) a weak design which expands the uniform seed into several separate bit strings (\textit{chunks}) that have limited overlap, in the sense that only a subset of bits from any two strings match, and (2) a 1-bit extractor which combines each of the chunks, in turn, with the weak {input} to produce one (near-)perfect bit. 
By applying the 1-bit extractor to each expanded chunk, the Trevisan extractor produces multiple bits of near-perfect randomness. 
Its drawback is its computation time, which often prohibits its use in practice. 
The implementation, based on the building blocks of \cite{mauerer2012modular}, in computation time $O(n^2 \log (\log (\frac{n^2}{\epsilon})))$, can be found in \Cref{subsec:trevian-imp} and has a seed {length}, $d$, of
\begin{align} 
    d &= at^2,
\end{align} with
\begin{align}
    t &\in \mathbb{P}_{\geq q}\quad : \quad q = 
    2 \lceil \log_2(n) + 2 \log_2(2m/\epsilon) \rceil, \\
    a &= \max \Big\{ 1 , \Bigl\lceil \frac{\log_2(m-2 \exp(1)) - \log_2(t-2 \exp(1))}{\log_2(2 \exp(1)) - \log_2(2 \exp(1)-1)} \Bigr \rceil \Big\},
\end{align} where $\mathbb{P}_{\geq q}$ denotes the set of primes that are larger or equal to $q$ and $\exp(1) \approx 2.718$ is the basis of the natural logarithm.
As for $\Circulant$ and $\Toeplitz$, $\Trevisan$ is directly quantum-proof \cite{de2012trevisan}, hence we state the claims as one theorem only.

\begin{theorem}[Combining parameters \cite{mauerer2012modular}, with quantum-proof in \cite{de2012trevisan}]
	The $\Trevisan$ extractor is a strong (classical-proof or) quantum-proof $(n_1 = n, k_1, n_2 = d, k_2= d, m, \epsilon)$-seeded randomness extractor, with
	\begin{align} 
		m \leq k_1 + 4 \log_2(\epsilon) - 4 \log_2(m)- 6.
	\end{align}
 \end{theorem}

Using the two-weak-source extension (\Cref{lemma:2-weak-source}), the $\Trevisan$ extractor is a classical-proof $(n_1 = n, k_1, n_2 = d, k_2, m, \epsilon)$-two-source randomness extractor strong in the {weak seed (with min-entropy $k_2$)}, where
	\begin{align} 
		m \leq k_1 + 4k_2 - 4d + 4 \log_2(\epsilon) - 4 \log_2(m) - 6, 
	\end{align}
 and, {using \Cref{Markov2}, a quantum-proof $(n_1 = n, k_1, n_2 = d, k_2, m, \epsilon)$-two-source randomness extractor in the Markov model with
 \begin{align}\label{params:trevisan-2-mm}
		m \leq \frac{1}{10} \left( k_1 + 4(k_2 - d) - 4 \log_2(m) + 18 \log_2 \epsilon + 9 \log_2 \left( \frac{4}{3} \right) - 6\right).
 \end{align}}

\noindent \textit{Trevisan with shorter seed length --} Our implementation uses the Block Weak Design from \cite{mauerer2012modular} which iteratively calls the weak design of \cite{hartman2003distribution} to generate the necessary chunks.
Using other weak designs allow for the seed length to be made shorter in practice, at the expense of increasing the entropy loss. 
For example, one can use our implementation with the weak design of \cite{hartman2003distribution} directly, to achieve a seed length $d = t^2$ at the expense of reducing the output by a factor of $m \rightarrow m/2\exp(1)$ i.e.\ giving a minimum entropy loss of at least $1 - 1/2\exp(1) \approx 82\%$\@.

\subsection{Von Neumann}
\label{sec:vn}
The $\VonNeumann$ deterministic extractor is a function $\VonNeumann(x): \{0,1\}^n \to \{0,1\}^m$ taking a single input $x = x_0, x_1, \ldots x_{n-1} \in \{0,1\}^n$ where the random variables that generate each bit form an \textit{exchangeable sequence}. 
More concretely, the $\VonNeumann$ extractor requires that the input probability distribution of $X$ generating the input $x$ satisfies
\begin{align}\label{eq:2} 
	\Pr(X_{2i} = 0 |E) = \Pr(X_{2i+1} = 0|E) = p_{i},
	\end{align}  
{for all $i \in \{0, \ldots, \lfloor n/2 \rfloor-1\}$, some $p_{i} \in [0,1]$ and some adversary side information $E$ (or alternatively, for a quantum-proof extractor, side information $\E$).}
Let $j = 0, \ldots, \lfloor n/2 \rfloor - 1$, the $j$th output bit of the $\VonNeumann$ extractor is then given by
\begin{align}
    \VonNeumann(\mathbf{x})_j = \begin{cases}
    x_{2j} &\quad \text{if} \hspace{0.5cm}  x_{2j} \neq x_{2j+1}\\
    \emptyset &\quad \text{otherwise} 
    \end{cases},
\end{align} where the empty set $\emptyset$ denotes that there is no {output bit} in that round.
Note that, contrary to the other constructions, the output {length} of the $\VonNeumann$ extractor is probabilistic. 
For example, a valid input $X$ is one that is IID with an unknown bias $p_i = p$ for all $i$\@. 
In this case, the approximate (due to finite sampling from $X$) output length is given by 
\begin{align} 
	m \approx p(1-p)n, 
\end{align} 
which gives an entropy loss $k - m$ of $(1 + (1-p)p/\log_2(\max\{p,1-p\}))k_1 \geq 3k_1/10$ i.e.\ implying {at least} $30\%$ entropy loss.  
This is significantly worse than the entropy loss when using the same input to an optimal seeded extractor, where $m = \lfloor -n\log_2(\max\{p, 1-p\}) - 2 \log_2(\frac{1}{\epsilon}) \rfloor$ i.e.\ approximately $0\%$ entropy loss when $\epsilon$ is constant.
Other extractors based on this construction, but optimising the output {length} by recycling unused bits, allow for this entropy loss to be improved, including the Elias and Peres extractor \cite{prasitsupparote2018numerical} and the generalised Von Neumann extractor \cite{gravel2021generalization}\@.
However, they are still unable to achieve the same entropy loss that is possible with seeded extractors, in general.
For completeness, we provide a pseudo-code implementation in \Cref{subsec:vn-imp} with {computation time} $O(n)$\@. 

\section{Cryptomite: Examples with code}
\label{sec:showcase}
In this section we showcase \texttt{Cryptomite} by giving code examples for privacy amplification in quantum key distribution (QKD) and randomness extraction in random number generation (RNG)\@. 
We give the {Python} code specific to the experimental demonstrations in \cite{jain2022practical, avesani2021semi}, as well as extensions that improve their results by relaxing assumptions (hence increasing security) and reducing resources. We note that one immediate benefit of using our extractors is also the numerical precision obtained by using the NTT for performance.
These concrete examples and code can easily be adapted to any setup requiring a randomness extractor.
More examples can also be found in the documentation.\\

{Remark that all the examples that we give in this section also showcase the limitation of the Trevisan extractor (and others) in practice. Indeed, for these examples the input lengths are too long for implementations that do not have quasi-linear computation time $O(n\log(n))$, i.e. that the usual notion of polynomial efficiency is irrelevant in practical quantum cryptography. We refer the reader to \Cref{fig:performance} for a numerical example of the performance with different input lengths.}

\subsection{\Cryptomite \hspace{0.1cm} for privacy amplification in QKD} \label{subsec:qkd}
In QKD, the goal is for two parties, Alice and Bob, to generate a shared secret key that is unpredictable to an adversary. 
After rounds of quantum and classical communication, Alice and Bob share an identical shared raw key which is only partially secret to the adversary (i.e.\ after state sharing and measurement, sifting, parameter estimation and error reconciliation)\@. 
Randomness extractors are used for privacy amplification, a subroutine which transforms the raw key (that has some min-entropy) into a final secret key that is (almost) completely secret to adversary (i.e.\ $\epsilon$-perfectly random to the adversary)\@. 
In standard protocols, this task is performed using a strong seeded extractor, which requires both Alice and Bob to have a shared (near-)perfect seed. \\

We consider the {continuous-variable} QKD demonstration of \cite{jain2022practical}, which aims at security against quantum adversaries and therefore requires a quantum-proof extractor.
From the experiment, Alice and Bob obtain $n = 1.738 \times 10^9$ bits of shared raw key as input for privacy amplification\footnote{Since the extractor input length is bigger than $2^{29}$, the extractor implementation will use the \texttt{bigNTT}, implying a throughput of $3-4\times$ less than that shown in \Cref{sec:performance}.}.
After evaluating the total min-entropy of the raw key, the authors compute the final secret key length of $m=41378264$ bits based on their given security parameters (including an extraction error of $\epsilon=10^{-10}$) when using their $\Toeplitz$ extractor implementation.
This extraction requires a perfect seed of length $d = m+n-1 = 1.738 \times 10^9 + 41378263$ and can be performed in a few lines of Python code using $\Cryptomite$, as shown in \Cref{fig:pa}\@.\\

\noindent \textit{Extensions -- }
The $\Circulant$ extractor can be used to generate the same amount of shared secret key, whilst substantially reducing the {length} of the seed. The code for this is given in \Cref{fig:pa-improv2}\@. 
The setup can be further improved, as in the experiment a quantum-RNG (QRNG) is used to generate the extractor seed \cite{gabriel2010generator} and its quality relies on the correct characterisation and modelling of the components in the device (an assumption which was shown to be potentially problematic in \cite{thewes2019eavesdropping})\@. 
We show how to relax this assumption by assuming that the min-entropy rate of the QRNG output is only $r = 0.99$, i.e.\ $k_2=r \cdot n_2$, and adjusting the output length of the extractor accordingly (in the quantum-proof product source model), see \Cref{fig:Table}\@. The code to perform the parameter calculation is given in \Cref{fig:pa-improv2}, but it could also easily be performed using the calculation module \texttt{from\_params} (see documentation)\@.

\begin{figure}[H] 
\begin{minipage}{1\textwidth}
  \begin{minted}{python}
  import cryptomite
  def privacy_amplification(raw_key_bits: list, seed_bits: list,
                            n_1 = 1.738 * 10**9, m = 41378264)
      """ Perform privacy amplification for the QKD protocol.

      Parameters
      ----------
      raw_key_bits : list of bits, derived from the measurement 
        outcomes (after sifting, error correction and parameter 
        estimation). 
      seed_bits : list of bits, generated independently.
      n_1: integer, the length of the raw key bit string.
      m: integer, the length of the output secret key bit string.

      Returns
      ---------
      list of bits, 
        the extracted output (i.e. the shared secret key).
      """
      # Initialise the Toeplitz extractor with the appropriate parameters:
      toeplitz = cryptomite.Toeplitz(n_1, m)
      # Perform Toeplitz extraction and return the output:
      return toeplitz.extract(raw_key_bits, seed_bits)
  \end{minted}
  \end{minipage}
    \caption{
  The implementation of privacy amplification for \cite{jain2022practical} using the $\Toeplitz$ extractor from \Cryptomite\@. \label{fig:pa}}
  \end{figure}

\begin{figure}[H] 
\begin{minipage}{1\textwidth}
  \begin{minted}{python}
  import cryptomite
  from math import floor, log2
  def improved_privacy_amplification(raw_key_bits: list, seed_bits: list,
                                     n_1 = 1.738 * 10**9, epsilon = 10**(-10), 
                                     m = 41378264, r = 0.99)
      """ Perform improved privacy amplification for the QKD protocol.

      Parameters
      ----------
      raw_key_bits : list of bits, derived from the measurement outcomes
        (after sifting, error correction and parameter estimation). 
      seed_bits : list of bits, generated independently.
      n_1: integer, the length of the raw key bit string.
      epsilon: float, the extractor error.
      m: integer, the length of the output secret key bit string.
      r: float, the min-entropy rate of the extractor (now weak) seed.
      
      Returns
      ---------
      list of bits, the extracted output (i.e. the shared secret key).
      """
      # Calculate the input min-entropy from the given extractor
      # output length and extractor error:
      k_1 = m - 2*log2(epsilon)
      # To use Circulant, the seed length needs to be a prime with 
      # primitive root 2, we find the one larger than or equal to 
      # n_1 + 1 (see Def. 9):
      n_adjusted = cryptomite.utils.next_na_set(n_1 + 1)
      # If n_adjusted > n_1 + 1, we increase the length of
      # the raw key to n_adjusted - 1 by padding 0's to it:
      if n_adjusted > n_1 + 1:
          raw_key_bits += [0]*(n_adjusted - n_1 - 1)
      # Take the correct number of seed bits (less than with Toeplitz):
      seed_bits = seed_bits[:n_adjusted]
      # Compute the new weak seed's min-entropy, given its 
      # min-entropy rate r:
      k_2 = r * n_adjusted
      # Compute the extractor's output length (product source 
      # model, see Fig. 3):
      m = floor(k_1 + k_2 - n_adjusted + 2*log2(epsilon))
      # Initialise the Circulant extractor with the appropriate parameters:
      circulant = cryptomite.Circulant(n_adjusted - 1, m)
      # Perform extraction and return the output:
      return circulant.extract(raw_key_bits, seed_bits)
  \end{minted}
  \end{minipage}
  \caption{
  Improved implementation of privacy amplification for \cite{jain2022practical} using the $\Circulant$ extractor with a weak seed. \label{fig:pa-improv2}}
  \end{figure}

\subsection{\Cryptomite \hspace{0.1cm} for randomness extraction in (Q)RNG} \label{subsec:qrng}
In random number generation (RNG), randomness extraction is used to process the raw output from an entropy source (sometimes called the noise source) into the final output that is $\epsilon$-perfectly random -- a process analogous to \textit{conditioning}
in the NIST standards \cite{turan2018recommendation}\@. 
In most protocols for quantum RNG (QRNG), this is done using (strong) seeded randomness extractors. 
To showcase $\Cryptomite$ for RNG, we replicate and extend the randomness extraction step of the {semi-device-independent} QRNG based on heterodyne detection \cite{avesani2021semi}, again giving code examples. \\ 

In \cite{avesani2021semi}, the randomness extractor error is chosen to be $\epsilon=10^{-10}$ and the length of the extractor input is $n_1 = 6.5 \times 10^{9}$ bits\footnote{Again, since the extractor input length is bigger than $2^{29}$, the extractor implementation will use the \texttt{bigNTT}, implying a throughput of $3-4\times$ less than that shown in \Cref{sec:performance}.}.
The experimental results and calculations certify a min-entropy rate of $k_1/n_1=0.08943$ and, therefore, a total min-entropy of $k_1 = \lfloor 0.08943 n_1 \rfloor = 581295000$ for the extractor input. 
The randomness extraction step is performed using the $\Toeplitz$ extractor, which gives an output length of $m = 581294933$ bits and requires a seed length of $n_2 = n_1 + m - 1 = 7081294932$ bits. 
This can be performed in a few lines of code using $\Cryptomite$, as shown in \Cref{fig:re}\@.\\

\noindent \textit{Extension --} As in the QKD example, the RNG case can be improved using other options given in our extractor library, reducing the seed length and allowing the option to relax the requirement on the seed min-entropy rate.
Using our $\Circulant$ extractor instead of the $\Toeplitz$ extractor saves $581294863$ seed bits. To implement this, the input length $n_1 = 6.5 \times 10^9$ must be adjusted to $n_1'$ such that $n_1' + 1$ is a prime number with primitive root 2. Based on this adjustment to $n_1$, the min-entropy must also be recalculated (using either \Cref{thm:short-seed-ext} or \Cref{lemma:shorten-entropy}). We find that the closest prime with primitive root 2 is $6.5 \times 10^9 + 69$, so we set $n_1' = 6.5 \times 10^9 + 68$ and pad the weak input with $68$ fixed bits.
Moreover, using a seeded extractor for random number generation demands a (near-)perfectly random seed as the resource, which leads to a circularity. 
As in the QKD example, using \texttt{Cryptomite} we introduce a min-entropy rate parameter $r$ that allows the extractor seed to be weakly random only, whilst still outputting near-perfect randomness.
The code for this extension can be found in \Cref{fig:re-dodis}\@.

\begin{figure}[H]
\begin{minipage}{1\textwidth}
  \begin{minted}{python}
  import cryptomite
  from math import floor, log2
  def randomness_extraction(raw_randomness: list, seed_bits: list, 
                            n_1 = 6500000000, epsilon = 10**-10,
                            k_1 = 581295000)
      """ Perform randomness extraction for the RNG protocol. 

      Parameters
      ----------
      raw_randomness : list of bits, the outcomes of measurements.
      seed_bits : list of bits, generated independently.
      n_1: integer, the length of the raw randomness bit string.
      epsilon: float, the extractor error.
      k_1: float, the total min-entropy of the raw randomness. 

      Returns
      ---------
      list of bits,
        the extracted output (i.e. the near-perfect randomness).
      """
      # Compute the extractor output length:
      m = floor(k_1 + 2*log2(epsilon))
      # Initialise the Toeplitz extractor with the appropriate
      # parameters:
      toeplitz = cryptomite.Toeplitz(n_1, m)
      # Perform Toeplitz extraction and return the output:
      return toeplitz.extract(raw_randomness, seed_bits)
  \end{minted}
  \end{minipage}
  \caption{
  The implementation of randomness extraction for \cite{avesani2021semi} using the $\Toeplitz$ extractor in \texttt{Cryptomite}\@. \label{fig:re}}
\end{figure}

\begin{figure}[H]
\begin{minipage}{1\textwidth}
  \begin{minted}{python}
  import cryptomite
  from math import floor, log2
  def improved_randomness_extraction(raw_randomness: list, seed_bits: list,
                                     n_1 = 6500000000, epsilon = 10**-10, 
                                     k_1 = 581295000, r = 0.99)
      """ Perform improved randomness extraction for the RNG protocol. 

      Parameters
      ----------
      raw_randomness : list of bits, the outcomes of measurements.
      seed_bits : list of bits, generated independently.
      n_1: integer, the length of the raw randomness bit string.
      epsilon: float, the extractor error.
      k_1: float, the total min-entropy of the raw randomness. 
      r: float, the min-entropy rate of the extractor (now weak) seed. 
      
      Returns
      ---------
      list of bits, 
        the extracted output (i.e. the near-perfect randomness).
      """
      # To use Circulant, the seed length needs to be a prime
      # with primitive root 2, we find the one closest 
      # to n_1 + 1 (see Def. 9):
      n_adjusted = cryptomite.utils.closest_na_set(n_1+1)
      # If n_adjusted < n_1 + 1, reduce the length of the raw
      # randomness to n_adjusted and reduce k_1 accordingly:
      if n_adjusted < n_1 + 1:
          k_1 -= n_1 + 1 - n_adjusted
          raw_randomness = raw_randomness[:(n_adjusted - 1)]
      # If n_adjusted > n_1 + 1, increase the length of
      # the raw randomness to n_adjusted by padding 0's:
      else: 
          raw_randomness += [0]*(n_adjusted - n_1 - 1)
      # Take the correct number of seed bits (less than with Toeplitz):
      seed_bits = seed_bits[:n_adjusted]
      # Compute the new weak seed's min-entropy, given its 
      # min-entropy rate r:
      k_2 = r * n_adjusted
      # Compute the extractor's output length (product source
      # model, see Fig. 3):
      m = floor(k_1 + k_2 - n_adjusted + 2*log2(epsilon))
      # Initialise the Circulant extractor with the appropriate parameters:
      circulant = cryptomite.Circulant(n_adjusted - 1, m)
      # Perform Circulant extraction and return the output:
      return circulant.extract(raw_randomness, seed_bits)
  \end{minted}
  \end{minipage}
  \caption{
  Improved implementation of randomness extraction for \cite{avesani2021semi} using the $\Circulant$ extractor with a weak seed. \label{fig:re-dodis}}
\end{figure}

\section{Conclusion and future work}
We have presented $\Cryptomite$, a software library of efficient randomness extractor implementations suitable for a wide range of applications. 
We made it easy to use and provided extensive documentation and examples to ensure that our extractors can be appended in a simple manner to any protocols.
The capacity of our extractors to tolerate large input lengths efficiently, whilst being numerically precise, makes it useful even for \mbox{(semi-)device} independent quantum cryptography (e.g.\ \cite{zhang2022device, bierhorst2018experimentally, avesani2021semi, jain2022practical, shalm2021device, zhang2021simple, liu2021device, li2021experimental, nadlinger2022experimental})\@.
We hope that our work helps to simplify the process of choosing the appropriate extractor and parameters with sufficient flexibility. \\ 

Finally, we list some future work and open questions.
\begin{itemize}
    \item We are in the process of implementing other randomness extractors, such as an efficient (in quasi-linear {computation time}) version of the Raz' extractor \cite{raz2005extractors} (\cite{foreman2024raz}) and the ones of Hayashi et al.\ \cite{hayashi-tsurumaru}. 
    These may be released in subsequent versions of $\Cryptomite$\@.
    \item The $\Trevisan$ extractor boasts an asymptotically small seed length. However, in practice it has the drawback that its computation time is large and, for small input lengths, the seed length is often longer than the input\footnote{{We remark that the seed length can be made small by reducing the output length, i.e. at the cost of the protocol's efficiency. See the exact statements of the seed length in \Cref{sec:trevisan}}}. Some interesting future work would be making a GPU implementation of $\Trevisan$, so that the one-bit extraction step is parallelised and fast in practice. One could alternatively find a combinatorial weak design that allows for a short seed length for small input lengths, whilst retaining minimal overlap between the chunks (see \Cref{sec:trevisan})\@.
    \item In order to implement the $\Toeplitz$ extractor in quasi-linear computation time, the Toeplitz matrix is generated from the (weak) seed which then gets embedded into a larger {circulant} matrix. It would be interesting to see if there exists a deeper link between the $\Circulant$ and $\Toeplitz$ (whilst $\Circulant$ is an extension of $\Dodis$)\@.
    \item To prove that some of our extractors are quantum-proof in the product source model, we employ the proof techniques of \cite{berta2021robust} and references therein which rely on obtaining a bound on the collision entropy of the extractor with a uniform seed.
    This bound is well understood for extractors based on two-universal hashing families. 
    It would be interesting to find a generalised theorem that allows for the security of any two-source extractor in the quantum-proof product source model (through a bound on the collision entropy of a general extractor or otherwise)\@.
\end{itemize}

{Since the completion of this work, we have been made aware of \cite{private-comms}, in which the authors' prove that the family of $\Dodis$ extractors is quantum-proof in the Markov model with substantially better parameters than those obtained by the generic reduction of \cite{markov-extractors}.}

\section{Acknowledgements}
We thank Kevin Milner and Kieran Wilkinson for valuable feedback and comments, Sean Burton for reviewing and improving the $\Cryptomite$ code, Dan Neville for a useful code review of our Trevisan extractor, Sherilyn Wright for designing the repository logo, Ela Lee and Matty Hoban for testing the first version of $\Cryptomite$, and Lluis Masanes for useful discussions. We also thank Martin Sandfuchs for pointing out that the generalisation to the Markov model (\Cref{Markov2} in \cite{markov-extractors}) is unnecessary for extractors that are already quantum-proof in the product source model, enabling us to derive \Cref{MarkovFor2Uni}, and Christoph Pacher for pointing out a mistake in the original proof of \Cref{thm-2-univ}.

\newpage
% \printbibliography
% \normalsize
% \bibliographystyle{unsrt}
% \bibliographystyle{plainurl}
\bibliographystyle{unsrtnat}
\bibliography{library.bib}
\appendix

\section{Appendices}

\subsection{Comparison of seeded extractors}
\label{sec:comparison_seeded}
Seeded extractors are used for numerous applications. Trevisan's extractor has the asymptotically shortest seed length of the seeded extractors implemented in this work. However, this is not always the case in practise. In this subsection, we provide a figure demonstrating that the asymptotic performance is not always preserved for short input lengths. We consider only those that have quantum-proof in the seeded case with near minimal entropy loss (i.e., not $\Dodis$).
\begin{figure}[H]
  	\centering
  	\includegraphics[width=0.6\textwidth]{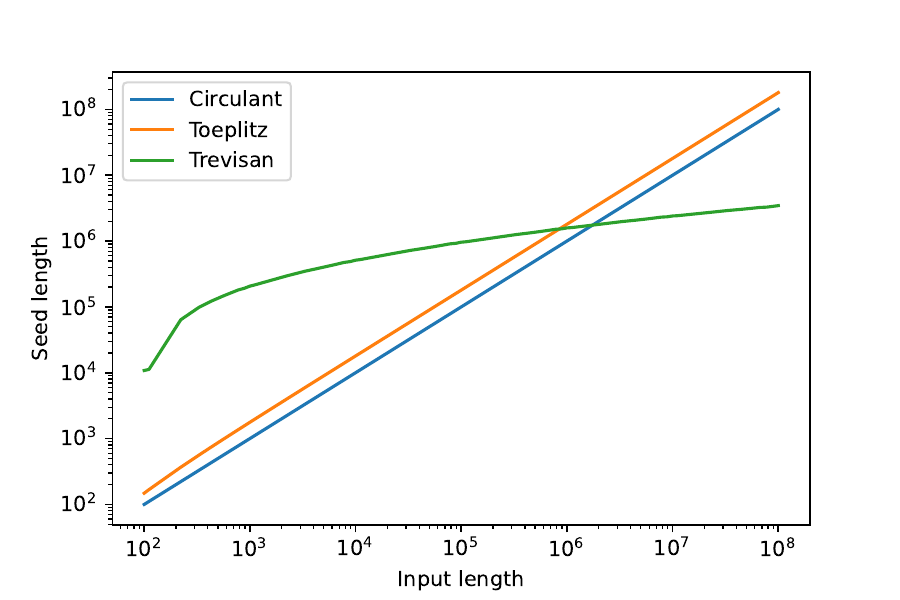}
  	\caption{{A plot displaying the seed length of our extractor algorithms as a function of the input length ($n_1$) for each seeded extractor of $\Cryptomite$. The seed length is calculated by fixing the input min-entropy to $k_1 =\frac{n_1}{2}$ and calculating the maximum output length according to \Cref{fig:Table}\@.} \label{fig:seeds}}
\end{figure}
We remark that the seed length of the $\Trevisan$ extractor depends on the output length $m$ and the extractor error $\epsilon$. For small $m$, the seed will be shorter -- a situation that naturally occurs in the case of $k_1 \ll n_1$ and may be relevant in certain experimental situations. Alternatively, the seed length can be further reduced by using a different configuration for the \textit{weak design}, at the cost of additional entropy loss. For a more detailed explanation, see \Cref{sec:trevisan}.

\subsection{The {number-theoretic} transform and the convolution theorem}
\label{subsec:ntt}

The $\Circulant$, $\Dodis$ and $\Toeplitz$ extractors in $\Cryptomite$ can be expressed as convolutions, allowing an efficient implementation by using the fast Fourier transform (FFT) or number-theoretic Transform (NTT). 
However, the FFT uses floating point arithmetic, which may cause loss of precision that is unacceptable, for example, in cryptographic applications. 
Instead, we use the NTT with modular arithmetic to implement the same ring operations without precision issues.

\begin{definition}[{Number-theoretic} transform (NTT)]\label{def:ntt} The {number-theoretic} transform on a vector $v$ of length $n$ is a discrete Fourier transform computed using the ring of integers modulo $p$ ($\mathbb{Z}/p\mathbb{Z}$) instead of the ring of complex numbers ($\mathbb{C}$) with primitive root $\omega$ (instead of $e^{-i\frac{2\pi}{n}}$) such that $\omega^n = 1$. The inverse NTT, $\NTT^{-1}$, is defined such that $\NTT^{-1}(\NTT(v)) = v$ and
\begin{align} \NTT(v)_i = \sum_{j=0}^{n-1}{v_j \cdot \omega^{jk}} \qquad
     \NTT^{-1}(v)_i = \sum_{j=0}^{n-1}{v_j \cdot \omega^{-jk}}
  \qquad \text{for $i=0,\ldots, n-1$} .\end{align}
\end{definition}

\begin{theorem}[Convolution theorem] \label{thm:convolution} The NTT of the circular convolution of two vectors $v$ and $w$ of length $n$ is the element wise product of the NTTs of $v$ and $w$:
\begin{align} \NTT(v * w)_i = \NTT(v)_i \cdot \NTT(w)_i \qquad \text{for $i=0,\ldots, n-1$} \end{align}
\end{theorem}
\begin{proof}
  {We modify the usual proof of the discrete convolution theorem to use $\mathbb{Z}/p\mathbb{Z}$ instead of $\mathbb{C}$.
  \begin{align}
    \NTT(v * w)_i
    &= \sum_k^{n-1}\left(\sum_{j=0}^{n-1}{v_j \cdot w_{k-j}}\right) \omega^{ik}\\
    &= \sum_{j=0}^{n-1}{v_j\omega^{ij}\sum_{k=0}^{n-1}{w_{k-j}\omega^{i(k-j)}}} \\
    &= \left(\sum_{j=0}^{n-1}{v_j\omega^{ij}}\right)\left(\sum_{t=0}^{n-1}{w_t\omega^{it}}\right) \qquad \text{Let $t = k-j$}\\
    &= \NTT(v)_i \cdot \NTT(w)_i
  \end{align}}
\end{proof}

The convolution theorem allows for the implementation of convolution in $O(n\log n)$ time using the NTT (or FFT), rather than the naive method requiring $O(n^2)$\@. This efficiency improvement makes randomness extraction feasible for {input lengths} $n > 10^6$, as shown in \Cref{sec:performance} ``Performance''\@.

\subsection{Proofs for the $\Circulant$ extractor}
\label{CirculantConstruction}
To prove that our $\Circulant$ construction is an extractor with the parameters in \Cref{sec:library-in-detail}, we first show that the family of functions derived from circulant matrices forms a two-universal hash family. With this established, we apply the Leftover Hashing Lemma \cite{impagliazzo1989pseudo} and Corollary 5.6.1 of \cite{renner2008security} to conclude that the construction is quantum-proof. The main difficulty in proving the two-universality of the $\Circulant$ extractor comes from certain `problematic' input pairs (those differing in every position). To resolve this, we pad the extractor input with a single constant bit, eliminating problematic cases while minimising additional seed length. 
We now state some useful Lemmas and Theorems.

\begin{lemma}[\cite{vazirani1987efficiency}] \label{circ-row-rank} Let $n$ be a prime with 2 as a primitive root modulo $n$. Let $d' \in \{0,1\}^{n} \setminus \{\{0\}^n, \{1\}^n\}$ be any binary vector of length $n$ that is not all zeros or all ones. Let $\mathrm{circ}(d') \in \{0,1\}^{n \times n}$ be the circulant matrix generated by $d'$, where the $i$th row, for $i \in \{0, \ldots, n-1\}$, is $d'$ right-shifted by $i$ positions. Then the rank of $\mathrm{circ}(d')$ over $\{0,1\}$ satisfies $\mathrm{rank}(\mathrm{circ}(d')) \geq n - 1$.
\end{lemma}

\begin{lemma}\label{lem:fullrank}
Let $C = \mathrm{circ}(d') \in \{0,1\}^{n \times n}$ be the circulant matrix generated by $d' \in \{0,1\}^{n}$, where $n$ is a prime such that 2 is a primitive root modulo $n$. Suppose $d' \notin \{\{0\}^n, \{1\}^n\}$. Then any submatrix of $C$ formed by selecting $m \leq n-1$ rows has full row rank over $\{0,1\}$.
\end{lemma}

\begin{proof}
By Lemma~\ref{circ-row-rank}, $C = \mathrm{circ}(d')$ has at-least rank $n-1$ over $\{0,1\}$. 
When $C$ has full rank, the Lemma is trivially true, so suppose that $C \in \{0,1\}^{n \times n}$ has $\mathrm{rank}(C) = n - 1$. In this case, the space spanned by the rows of $C$ has dimension $n-1$, implying there exists a unique (up to scaling) linear dependence among the $n$ rows. We can write this, representing the $i$th row as $r_i \in \{0,1\}^{n}$, as
\begin{align}
    \bigoplus_{i=0}^{n-1} \alpha_i r_i = \{0\}^n, \quad \text{for some } \boldsymbol{\alpha} = (\alpha_0, \ldots \alpha_{n-1}) \in \{0,1\}^{n},\ \boldsymbol{\alpha} \neq \{0\}^{n},
\end{align} where the elementwise row addition is done modulo 2.

For any matrix with rows $\{r_i\}_i$ and a linear dependency vector $\boldsymbol{\alpha}$, any permutation of the columns preserves the linear dependency. 
In the case of a circulant matrix, shifting each column to the right by $j$ positions (cyclically) is equivalent to shifting each row element, for all rows, to the right by $j$ positions, resulting in the row that was originally $j$ rows below. Specifically, for any $j \in \{0, \ldots, n-1\}$, the $i$th row after cyclically permuting the columns by $j$ positions becomes $r_{(i + j) \bmod n}$. This means that we have 
\begin{align}
    \bigoplus_{i=0}^{n-1} \alpha_i r_i = \bigoplus_{i=0}^{n-1} \alpha_{i} r_{(i + j) \bmod n} = \{0\}^n, 
\end{align} for any $j \in \{0, \ldots, n-1\}$. By relabelling the coefficients accordingly, this expression is equivalent to
\begin{align}
    \bigoplus_{i=0}^{n-1} \alpha_i r_i = \bigoplus_{i=0}^{n-1} \alpha_{(i-j) \bmod n} r_i = \{0\}^n.
\end{align}
The second term corresponds to a linear dependency given by $j$ left cyclic shifts of $\boldsymbol{\alpha}$, implying that any cyclic shift of a dependency $\boldsymbol{\alpha}$ is also a dependency.

Putting this together: (i) since $\mathrm{rank}(C) = n - 1$, the null space of $C$ is one-dimensional, so any nonzero vector $\boldsymbol{\alpha}$ in the null space is unique up to scalar multiplication; and (ii) since $C$ is circulant, any cyclic shift of a dependency vector is also a dependency. The only non-all-zero vector in $\{0,1\}^n$ that is invariant under all cyclic shifts and unique up to scalar multiplication is the all-ones vector, i.e., $\boldsymbol{\alpha} = (1, 1, \dots, 1)$.  Thus, the only linear dependency among the rows of $C$ is $\sum_{i=0}^{n-1} r_i = \{0\}^n$.

This dependency involves all $n$ rows, so any submatrix formed by selecting $m \leq n - 1$ rows must have full row rank (i.e., rank $m$) over $\{0,1\}$. This completes the proof.
\end{proof}

\begin{lemma} \label{lem:short-prob}
Let $Y \in \{0,1\}^{n}$ be uniformly distributed, and let $G \in \{0,1\}^{m \times n}$ with $m \leq n - 1$. Then,
\begin{align}
    \Pr(GY = \{0\}^m) = \frac{|\ker(G)|}{2^{n}}.
\end{align}
\end{lemma}

\begin{proof}
Since $Y$ is uniformly distributed over $\{0,1\}^n$, each vector $y \in \{0,1\}^n$ occurs with probability $2^{-n}$. The event $GY = \{0\}^m$ occurs precisely when $Y \in \ker (G)$, which is a subspace of $\{0,1\}^n$. Thus,
\begin{align}
    \Pr(GY = \{0\}^m) = \Pr(Y \in \ker(G)) = \frac{|\ker(G)|}{2^n}.
\end{align}
\end{proof}

\begin{theorem} \label{thm-2-univ}
	The function defined by the Circulant extractor is a two-universal family of hash functions.
\end{theorem}

\begin{proof}
    The function $\Circulant$ forms a two-universal family of hash functions if, for all distinct $x, \tilde{x} \in \{0,1\}^{n-1}$ and for $Y \in \{0,1\}^n$ uniformly distributed (i.e., $p_Y(y) = 2^{-n}$ for all $y$), 
    \begin{align} \label{2-univ}
        \Pr\left(\Circulant(x, Y) = \Circulant(\tilde{x}, Y)\right) \leq 2^{-m}
    \end{align}
    holds for any $m \leq n - 1$.
    
   Define $x' = [x, 0]$ and $\tilde{x}' = [\tilde{x}, 0]$. Let $d' = x' \oplus \tilde{x}'$, where $\oplus$ denotes bitwise addition modulo 2, and $C_m$ be the $m \times n$ submatrix consisting of the first $m$ rows of $\mathrm{circ}(d')$. Using this notation, the collision probability can be rewritten as
    \begin{align}
        \Pr\left(\Circulant(x, Y) = \Circulant(\tilde{x}, Y)\right) 
        &= \Pr(C_mY = \{0\}^m) = p_{C_mY}(\{0\}^m).
    \end{align}
    Next, $x$ and $\tilde{x}$ differ in at least one element by definition, so $d' \neq \{0\}^n$. Moreover, since $x'$ and $\tilde{x}'$ share the same last bit due to padding, we also have $d' \neq \{1\}^n$. Therefore, by Lemma~\ref{lem:fullrank}, the submatrix $C_m$ has rank $m$ over $\{0,1\}$ for any $m \leq n-1$.

    By the rank-nullity theorem, if a matrix has rank $m$ over $\{0,1\}$, then its kernel has dimension $n - m$, and hence size $2^{n - m}$.  
    Therefore, since $C_m$ has rank $m$, its kernel has size $2^{n - m}$. Applying Lemma~\ref{lem:short-prob}, we obtain
    \begin{align}
        p_{C_mY}(\{0\}^m) = \frac{2^{n-m}}{2^n} = 2^{-m},
    \end{align}
    as required.
\end{proof}

Now, we have the ingredients to prove the main theorem, \Cref{circSEED}, which follows directly from \Cref{thm-2-univ} and Corollary 5.6.1 from \cite{renner2008security}\@.

\subsection{$\Dodis$ (and $\Circulant$) implementation}
\label{subsec:dodis-imp}
Concretely, our implementation of the $\Dodis$ extractor follows \cite{foreman2020practical} with $\{A_i\}_i$ chosen to be right cyclic shift matrices, which means {losing} one output bit compared the optimal choice for the extractor. 
We also require that $n$, the input length, is prime with primitive root 2 and $X = x \neq \{0\}^{n}, \{1\}^n$. The set of matrices {$A_0, A_1 \ldots A_{n-1}$} are defined as the $n \times n$ matrices{
\begin{align}
	A_0 = \begin{bmatrix}
		1 & 0 & 0 & \ldots & \ldots & 0 \\
		0 & 1 & 0 & \ldots & \ldots & 0 \\
		0 & 0 & 1 & \ldots & \ldots & 0 \\
		\vdots & & \ddots & & \vdots  \\
		0 & 0 & 0 & \ldots & 1 & 0 \\
		0 & 0 & 0 & \ldots & 0 & 1 \\ 
	\end{bmatrix},    \quad A_1 = \begin{bmatrix}
		0 & 1 & 0 & \ldots & \ldots & 0 \\
		0 & 0 & 1 & \ldots & \ldots & 0 \\
		0 & 0 & 0 & \ldots & \ldots & 0 \\
		\vdots & & \ddots & & \vdots  \\
		0 & 0 & 0 & \ldots & 0 & 1 \\
		1 & 0 & 0 & \ldots & 0 & 0 \\ 
	\end{bmatrix}, \ldots,
\end{align}} and satisfy the constraint that the sum of any subset $B \subset \{A_i\}_i$ has rank {at least} $n-1$, as proved in \cite{vazirani1987efficiency}.
In this form, the extractor can be re-written as the matrix-vector multiplication 
\begin{align}
	\Dodis(x,y) = \left(\begin{bmatrix}
		x_0 & x_1 & x_2 & \ldots & \ldots & x_{n-1} \\
		x_{n-1}  & x_0 & x_1 & \ldots & \ldots & x_{n-2}  \\
		x_{n-2}  & x_{n-1}  & x_0 & \ldots & \ldots & x_{n-3}  \\
		\vdots & & \ddots & \ddots & & \vdots  \\
		x_2 & x_3 & x_4 & \ldots & x_0 & x_1 \\
		x_1 & x_2 & x_3 & \ldots & x_{n-1}  & x_0\\
	\end{bmatrix} 
	\begin{bmatrix}
		y_0 \\
		y_1 \\
		\vdots \\
		\vdots \\
		y_{n-1}
	\end{bmatrix}\right)_{{0:m-1}},
\end{align} where $x = x_0, x_1, \ldots x_{n-1}$, $y = y_0, y_1, \ldots y_{n-1}$ and the subscript {$0:m-1$} denotes the first $m$ elements (bits) of the matrix-vector multiplication. \\

Although this construction is not optimal in the sense that it loses one output bit compared to the optimal construction, it allows the overall function to be re-written as a circular convolution (\cite{foreman2020practical}, Appendix D.1, Definition 10)\@.
This means that we can write
\begin{align}
	\Dodis(x,y)_i &= \Bigl(\mathsf{R}(x) * y\Bigr)_i \\
	&=  \sum^{n-1}_{j = 0} \mathsf{R}(x)_{i-j} \cdot y_j,
\end{align} {where subscript $i \in \{0, \ldots m-1\}$ denotes the $i$th output, $\mathsf{R}(x) = x_0, x_{n-1}, x_{n-2}, \ldots, x_1$, and indices are mod $n$.}
The output consists of the first $m$ elements of this convolution. 
The extractor algorithm is given by the pseudo-code in \Cref{alg:dodis}\@. \\

\begin{algorithm}
	\caption{Implementation of the $\Dodis$ Extractor}\label{alg:dodis}
	\begin{algorithmic}[1]
		\Function{ExtDodis$_{m, n}$}{u, v}
            \State $L \leftarrow 2^{\text{floor}(\log_2(2n -2) +1)}$ \Comment{Calculate NTT size}
		\State \textbf{create} $u'$ of size $L$
		\State \textbf{create} $v'$ of size $L$
		\For {$i=0\:\ldots\: n-1$}
		\State $u'[i] \leftarrow u[(n - i) \mod n]$
		\State $v'[i] \leftarrow v[i]$
		\EndFor
		\State $w \leftarrow conv(u', v)$
		\State \textbf{return} first $m$ bits of $w$
		\EndFunction
	\end{algorithmic}
\end{algorithm}

\textbf{Total Computation Time:} The $\Dodis$ extractor requires two pre-processing steps of $O(n)$: first, to reduce $n$ so that it is a prime with 2 as a primitive root, and second, to check that the input $X$ is not the all $0$ or all $1$ string. 
The extractor itself can be implemented using the NTT, as shown in \Cref{alg:dodis}. As shown in \Cref{thm:convolution}, this has computation time $O(n \log n)$.
Therefore, the total computation time for implementing the $\Dodis$ extractor is $O(n \log n)$ (sometimes called near-linear or quasi-linear).  \\

\noindent \textit{Circulant extractor implementation -- } The $\Circulant$ extractor uses the $\Dodis$ implementation described above, with one minor difference that does not affect the overall computation time: the input $x$ is padded with a single $0$ bit. 

\subsection{$\Toeplitz$ implementation}
\label{subsec:toeplitz-imp}

Although the definition of the $\Toeplitz$ extractor is essentially a circular convolution, the
$O(n\log n)$ algorithm in \Cref{subsec:ntt} expects the input and output lengths of the convolution
to be equal, so it cannot be directly applied. 
Therefore, to implement $\Toeplitz$ in $O(n\log (n))$ time, we must embed the Toeplitz matrix from \Cref{eq:toep}, with $m$ rows and $n$ columns, within a square circulant matrix.
Concretely, the Toeplitz matrix is embedded in the top-left quadrant of the square $(n + m - 1) \times (n + m - 1)$ circulant matrix \Cref{toeplitz-embed}:
{
\begin{align} \label{toeplitz-embed}
\mathrm{circ}(y_0, y_{n + m - 2}, y_{n + m - 3}, \ldots, y_{1}) =
\begin{bmatrix}
    \mathrm{toep}(y) & \cdot \\
     \cdot & \cdot
\end{bmatrix} ,
\end{align}
where $\mathrm{circ}(\cdot)$ denotes the circulant matrix generating function as in \Cref{eq:circ-matrix} and $\mathrm{toep}(y)$ denotes the Toeplitz matrix from \Cref{eq:toep}. Then, by padding the weak input $x \in \{0,1\}^n$ with $m-1$ zeros (i.e., $x \to [x, \{0\}^{m-1}]$), the $\Toeplitz$ extractor can be recovered by the matrix-vector multiplication $(\mathrm{circ}(y_0, y_{n + m - 2}, y_{n + m - 3}, \ldots, y_{1}) x')_{0:m-1}$ where $x' = [x, \{0\}^{m-1}]$. Using this embedding, the $\Toeplitz$ extractor can be implemented using the NTT in $O(n\log n)$ computation time.}
The full implementation of the $\Toeplitz$ extractor can be seen in the pseudo-code in \Cref{alg:toeplitz}\@. \\

\begin{algorithm}
	\caption{Implementation of the $\Toeplitz$ Extractor}\label{alg:toeplitz}
	\begin{algorithmic}[1]
		\Function{ExtToeplitz$_{n, m}$}{u, v}
            \State $L \leftarrow 2^{\text{floor}(\log_2(2n) +1)}$ \Comment{Calculate NTT size}
		\State \textbf{create} $u'$ of size $L$
		\State \textbf{create} $v'$ of size $L$
		\For {$i=0\:\ldots\: m-1$}
		\State $v'[i] \leftarrow v[i]$
		\EndFor
		\For {$i=0\:\ldots\: n-1$}
		\State $u'[i] \leftarrow u[i]$
		\State $v'[L - n + 1 + i] = v[m + i]$
		\EndFor
		\State $w \leftarrow conv(u', v')$
		\State \textbf{return} first $m$ bits of $w$
		\EndFunction
	\end{algorithmic}
\end{algorithm}

\textbf{Total Computation Time:} The Toeplitz extractor requires no pre-processing steps and can be computed by the convolution implemented using the NTT, embedding the Toeplitz matrix in a larger {circulant} matrix.
This has total computation time $O(n \log n)$\@.

\subsection{$\Trevisan$ implementation}
\label{subsec:trevian-imp}
We follow the implementations by Mauerer et al.\ \cite{mauerer2012modular}, which combines multiple iterations of the weak design from Hartman and Raz \cite{hartman2003distribution} to create a \textit{block weak design}, then uses the strong {Reed-Solomon Hadamard (denoted $\mathsf{RSH}$)} polynomial hashing based one-bit extractor constructed from a combination of the Reed-Solomon and the Hadamard code (see \cite{mauerer2012modular}, `Section III.\ DERIVATIONS')\@. 
Informally, our implementation of the $\Trevisan$ extractor takes an input $x \in \{0,1\}^n$ and a seed $y \in \{0,1\}^d$ and extracts $m$ output bits by 
\begin{enumerate}
    \item Using the block weak design, generate $m$ sub-strings of length $t$ from $y$ that have maximum overlap $r=1$ (where $r$ is defined in \Cref{eq:overlap}).
    \item Using each of these $m$ sub-string to (respectively) seed a $\mathsf{RSH}$ one-bit one-bit extractor, {process} the extractor input $x$ into a near-perfect bit as output.
    \item Combine the $m$ single output bits from the individual $\mathsf{RSH}$ one-bit extractors.
\end{enumerate}
In order for the sub-string generation and one-bit extraction procedures to fit together, the implementation requires the extractor seed, $y$, to be of length $d=at$ (following the notation of \cite{mauerer2012modular}), where 
\begin{align}
    \label{eq: t} t &\in \mathbb{P}_{\geq q}\quad : \quad q = 
    2 \lceil \log_2(n) + 2 \log_2(2m/\epsilon) \rceil, \\
    \label{eq:a} a &= \max \{ 1 , \Bigl\lceil \frac{\log_2(m-2 \exp(1)) - \log_2(t-2 \exp(1))}{\log_2(2 \exp(1)) - \log_2(2 \exp(1)-1)} \Bigr \rceil\},
\end{align} where $\mathbb{P}_{\geq q}$ denotes the set of primes that are larger or equal to $q$ and $\exp(1) \approx 2.718$ is the basis of the natural logarithm.
For what follows in this section, we use $\epsilon_1$ to denote the per-bit error (coming from the one-bit extractor) and the total error is given by $\epsilon = m \epsilon_1$. 
We now formalise this description. \\

\begin{definition}[Weak design, from \cite{mauerer2012modular}] \label{def:weakdesign-definition}
	A family of sets $S_1, \ldots, S_m$, where $S_i \subset \{1,\ldots, d\}$ for all $i \in \{1, \ldots, m\}$ constitutes a $(m, t, r, d)$-weak design if
	\begin{itemize}
        \item Each set $S_i$ has the same length, $t$.
		\item All sets in the family $S_1, \ldots, S_m$ have maximum overlap $r$, i.e.\ satisfies the condition 
		\begin{align} \label{eq:overlap} \forall i: \quad \sum_{j=1}^{i-1} 2^{\left|S_j \cap S_i\right|} \leq r m . \end{align}
	\end{itemize}
\end{definition}
Each set $S_i$ can be understood as a set of $t$ indices/positions of a $d$ length bit string. 
For some given bit string $y$, we use $y_{S_i}$ to denote the sub-string of $y$, of length $t$, where the elements of $y_{S_i}$ are the elements of $y$ at indices $S_i$.

\begin{definition}[1-bit extractor] \label{def:onebitextractor}
	A 1-bit extractor is a strong seeded extractor $\mathsf{Ext}_s^1:\{0,1\}^n \times \{0,1\}^d \to \{0,1\}$, i.e.\ a strong seeded extractor with output length exactly 1. 
\end{definition} 

\noindent The $i$th output bit of the $\Trevisan$ extractor, $\Trevisan(x, y):\{0, 1\}^n \times \{0, 1\}^{d} \to \{0, 1\}^m$, can be written as:
\begin{align}  
\Trevisan(x, y)_i = \mathsf{Ext}_s^1(x, y_{S_i}) 
\end{align}
where $i \in \{ 1, \ldots, m \}$ denotes the $i$th output bit,  $S_1, \ldots, S_m$ are the outputs of the weak design.\\

\noindent \textbf{Block weak design:}
In our implementation, we follow that of \cite{mauerer2012modular} and use $a$ (\Cref{eq:a}) instances of the weak design from Hartman and Raz \cite{hartman2003distribution} to form a single \textit{block weak design}. 
Using the Hartman and Raz weak design obliges $t$ to be prime. 
The full details and parameters can be found directly in \cite{mauerer2012modular}, `Section III.\ B.\ Weak Designs'. \\

\noindent \textbf{$\mathsf{RSH}$ 1-bit extraction:} {After using the block weak design to split the seed into $m$ sub-strings, we perform $m$ $\mathsf{RSH}$ one-bit extractions, where the one bit extractor is constructed from the concatenation of a Reed-Solomon and a Hadamard code (see \cite{mauerer2012modular}, `Section III.\ C.\ 3.\ Polynomial hashing')\@.}
Each one-bit {extractor} takes as input the extractor input $x \in \{0,1\}^n$ and one of the $m$ sub-strings of the seed $y$ generated by the block weak design, $y_{S_i} \in \{0,1\}^t$ for some $i \in \{1, \ldots, m\}$\@.
To implement $\mathsf{RSH}$, the seed sub-string $y_{S_i}$ is further split into two sub-strings of $l$ bits, where the first is used in the Reed-Solomon step and the second in the Hadamard step.
The initial $\mathsf{RSH}$ seed is of length $t \geq 2l$, where $l$ is  
\begin{align} 
	l = \lceil\log_2(n) + 2\log_2(2/\epsilon_1)\rceil, 
\end{align} from \cite{mauerer2012modular}, `Section III.\ C.\ 3.\ Polynomial hashing'.
The first step in the $\mathsf{RSH}$ extractor is the Reed-Solomon step. 
This is implemented as follows:
\begin{itemize}
	\item Split the extractor input $x$ into $s=\lceil n/l \rceil$ chunks $x_i$. Pad the last chunk $x_s$ with $0$s if need be to make all chunks of length $l$.
	\item View each chunk $x_i$ as an individual element in the field of $\mathbb{GF}(2^l)$ i.e.\ the binary representation of it, for example, if $l=3$ and $x_0=\{0,0,1\}$ then $x_0=1$ in the field $\mathbb{GF}(2^3=8)$\@. 
	\item Then, we evaluate the polynomial \Cref{eq:poly} (calculated using finite field multiplication) for each of the $s$ elements $x_i$ in $\mathbb{GF}(2^l)$ constructed form the extractor input, where $\alpha_1 \in \mathbb{GF}(2^l)$ is the element generated by using the first $l$ seed bits (i.e.\ the first half) to fix each of the $l$ coefficients.
 \begin{align} \label{eq:poly}
     p_{\alpha_1}(x) = \sum_{i=1}^s x_i \alpha_1^{s-i}. 
 \end{align}
\end{itemize}
Next, we combine the Reed-Solomon step with a Hadamard Step:
\begin{itemize}
	\item Let $\alpha_2 \in \mathbb{GF}(2^l)$ be the finite field element generated using the second $l$ seed bits (i.e.\ the second half) to fix each of the $l$ coefficients. 
    \item Output $z = \oplus_{i=1}^{l} (\alpha_2)_i p_{\alpha_1}(x)_i$, where $\oplus$ denotes addition modulo 2.
\end{itemize} 

\noindent \textbf{Total $\Trevisan$ computation time:}
The combined computation time is
\begin{align} \label{eq:trev-comp1} O(\underbrace{m\log(m)t\log(t)\log(\log(t))}_{\text{weak design}} + \underbrace{msl\log(l)}_{\text{$m$ 1-bit extractions}}) \end{align} and we use the following relationships to determine the overall computation time in terms
of $n$ and $\epsilon_1$:
\begin{align}
n \approx m \quad s = \left\lceil \frac{n}{l} \right\rceil \quad
t \approx O\left(\log(n) + \log\left(\frac{2}{\epsilon_1}\right)\right) = 
O\left(\log\left(\frac{n}{\epsilon_1}\right)\right) \quad 
l = \frac{t}{2} = O\left(\log\left(\frac{n}{\epsilon_1}\right)\right).
\end{align}
This allows us to re-write \Cref{eq:trev-comp1} as 
\begin{align}
	&O\left(
	n\log(n) \log\left(\frac{n}{\epsilon_1}\right)
	\log\log\left(\frac{n}{\epsilon_1}\right) \log\log\log\left(\frac{n}{\epsilon_1}\right) + 
	n^2\log\log\left(\frac{n}{\epsilon_1}\right)\right) \\
	&\approx \label{eq:trev-complex}
	O\left(
	\underbrace{n\log(n)\log\left(\frac{n}{\epsilon_1}\right)}_{\text{weak design}} +
	\underbrace{n^2\log\log\left(\frac{n}{\epsilon_1}\right)}_{\text{$m$ 1-bit extractions}}\right)
\end{align}
The weak design could be pre-computed to reduce the overall computation time of the Trevisan extraction, but the dominant term in \Cref{eq:trev-complex} comes from the $m$ 1-bit extractions.

\subsection{$\VonNeumann$ implementation}
\label{subsec:vn-imp}
The $\VonNeumann$ extractor is implemented in the following pseudo-code.

\begin{algorithm}[H]
	\caption{Implementation of the $\VonNeumann$ Extractor}\label{alg:vn}
	\begin{algorithmic}[1]
		\Function{ExtVonNeumann}{b} \\
                out = []
		\For {$i=0\:\ldots\: \lfloor \textbf{len}(b) / 2 \rfloor - 1$}
		\If{$b[2i] \neq b[2i+1]$}
		\State out.append($b[2i]$)
		\EndIf
		\EndFor
		\State \textbf{error} \Comment{Error if all bits are the same.}
		\EndFunction
	\end{algorithmic}
\end{algorithm}

\textbf{Total Computation Time:} Trivially, this extractor can be implemented with linear computation time ($O(n)$).

\end{document}